\documentclass[twocolumn,showpacs,preprintnumbers,
superscriptaddress,amsmath,floatfix,prd]{revtex4}

\usepackage{graphicx}

\renewcommand\a{\alpha}
\renewcommand\b{\beta}

\renewcommand\d{\delta}
\newcommand\e{\epsilon}

\newcommand\m{\mu}

\newcommand\G{\Gamma}

\newcommand{\non}{\nonumber\\}

\newcommand{\be}{\begin{equation}}
\newcommand{\ee}{\end{equation}}
\newcommand{\bea}{\begin{eqnarray}}
\newcommand{\eea}{\end{eqnarray}}
\newcommand{\ba}[1]{\begin{array}{#1}}
\newcommand{\ea}{\end{array}}

\begin{document}

\preprint{MIT-CTP-3722}

\title{Stability conditions and Fermi surface topologies in a superconductor} 

\author{Elena Gubankova}
\email{elena1@mit.edu}
\affiliation{Center for Theoretical Physics, Massachusetts Institute 
of Technology, Cambridge, MA 02139, USA}

\author{Andreas Schmitt}
\email{aschmitt@wuphys.wustl.edu}
\affiliation{Center for Theoretical Physics, Massachusetts Institute 
of Technology, Cambridge, MA 02139, USA}
\affiliation{Department of Physics, Washington University St Louis, MO, 63130, USA}

\author{Frank Wilczek}
\email{wilczek@mit.edu}
\affiliation{Center for Theoretical Physics, Massachusetts Institute 
of Technology, Cambridge, MA 02139, USA}

\date{\today}

\begin{abstract}
Candidate homogeneous, isotropic superfluid or superconducting states of paired fermion species with 
different chemical potentials, can lead to quasiparticle excitation energies that vanish at either 
zero, one, or two spheres in momentum space.  With no zeroes, we have 
a conventional BCS superconductor. The other two cases, ``gapless'' superconductors,
appear in mean field theory 
for sufficiently large mismatches and/or sufficiently large coupling strengths.  Here we examine several stability criteria for those candidate phases. 
Positivity of number susceptibility appears to provide the most powerful constraint, and renders
all the two-zero states that we have examined mechanically unstable. 
Our results should apply directly to ultracold fermionic atom systems.     
\end{abstract}


\maketitle

\section{Introduction}
\label{intro}

In a conventional superconductor the Fermi surface disappears \cite{bcs}.  Anisotropic 
fermion pairing is known, in several cases, to leave behind residual lower-dimensional effective Fermi surfaces: excitations of infinitesimal energy exist only in certain directions, defined by isolated points or lines in momentum space.  This situation occurs for instance in the {\it A} phase of helium-3 
\cite{helium3}, in high-temperature superconductors \cite{highTc} or in certain phases in 
color-superconducting quark matter \cite{spin1}.

It has been suggested that
isotropic multi-fermion systems with attraction between species that have Fermi surfaces of different size 
could lead to a more extreme version of this phenomenon, 
wherein pairing and superfluidity could coexist with gapless excitations across entire spheres.   Indeed, there is competition between the possibility to lower the energy by pairing one-particle states of equal and opposite momenta (${\bf p}$, $-{\bf p}$) coherently \cite{bcs}, to take advantage of the favorable interaction energy, and the fact that minimizing the kinetic energy does not put low-energy states at such momenta.   Thus plausibly it might become favorable to move fermions into different momentum states before pairing, taking a hit in kinetic energy, in order to profit from interaction energy.   
Depending on where
the excess fermions stayed, since it was too costly to move them, and on
the extent of the superfluid gap, ``scars'',
in the form of discontinuities in the occupation number, could be left 
behind.
Such scars appear as residual Fermi surfaces.

If the mismatch is small and the gap is large, any scars are covered over.
In that case the two fermion species
have equal number densities and the superfluid state
does not support gapless single-particle excitations.  If it is 
appropriate to
think that fermions of one species adjust to the other, e.g. by promoting
fermions from the top of the smaller Fermi surface to match the larger 
Fermi
surface and leaving the distribution with larger Fermi surface almost 
untouched,
then one free Fermi surface is left behind. 
This phase was proposed at large particle number asymmetry and strong coupling  
\cite{pao}. Here, the entire remaining Fermi sphere is occupied by unpaired particles of one species.
If it is appropriate to think that the compromise is reached 
by
moving some fermions up from the smaller Fermi surface and others down 
from the
larger Fermi surface while leaving unpaired fermions in between,
then two free surfaces are left behind.  This has been
termed the ``breached'' proposal \cite{breach}.  (In 
the
literature the terminology is not entirely uniform, but this usage seems
appropriate and convenient.) At large mass ratio, the dominant pairing 
comes
from the interior region \cite{Liu:2002gi}.

In a system with Fermi surface mismatch and $p$-wave pairing, the effective
residual Fermi surface can assume complicated two-dimensional shapes, e.g., topologically equivalent to 
a torus \cite{Gubankova:2004wt}.
In this paper, however, we consider only isotropic effective masses and interactions, and s-wave pairing.

Before describing the technical content of this paper, let us mention two concrete physical systems in which 
these unconventional superconducting or superfluid states are expected and possibly already observed. 

Remarkable control has been achieved over systems of ultracold fermionic atoms.  Many important parameters such as the number densities and the 
coupling strength can be varied.  These systems are studied in 
optical traps where an external magnetic field provides the ``knob'' to 
control the coupling strength around a Feshbach resonance \cite{exp}.  By choosing
different magnetic fields, the crossover from the BCS to the Bose-Einstein condensed (BEC) region 
can be observed. For theoretical studies on this crossover see for instance \cite{BECBCS}. Recently 
experimental studies have been extended to systems of atoms in two different spin states with
mismatched number densities \cite{KetterleImbalancedSpin}. What was a crossover 
in the symmetric situation then
appears to resolve into one or more phase transitions. For some values of the mismatch and the coupling, the results seem to suggest a mixed phase, i.e., 
a coexistence of spatially separated superfluid and normal phases \cite{HuletPhaseSeparation}, as 
predicted in Ref.\ \cite{Bedaque:2003hi}.

Another example for unconventional superconductivity is quark matter in the interior of neutron 
stars. If sufficiently cold and dense, this matter is color-superconducting \cite{Bailin:1983bm,reviews}. 
Quark matter can, in principle, form Cooper pairs in many 
different patterns.   At asymptotically high densities, where the quark masses can be neglected, the highly symmetric and fully gapped color-flavor locked (CFL) phase \cite{Alford:1998mk} is favored.  
At intermediate densities, many more pairing patterns are thinkable. All of them 
involve pairing of quarks with different Fermi surfaces \cite{Rajagopal:2005dg}.   
Gapless phases are conceivable, though the question of the true ground
state is currently unsolved.  While in this paper we focus on nonrelativistic two-fermion systems, and
do not analyze relativistic quark matter, very similar considerations arise in such an analysis.  Indeed, after completion of the work reported here, we learned of closely related work on relativistic systems \cite{igor}. 
   
In this paper we compute several stability criteria for these states within a simple model field theory. 
Specifically, we consider a $U(1)\times U(1)$ gauge theory, where each of the 
gauge fields couples to one of the fermions.  These fermions can have different chemical potentials.  
We work in the mean-field approximation and at zero temperature, but do not restrict ourselves to 
weak coupling.  We compute the electric and magnetic screening mass matrices from the 
zero-energy, low-momentum limit of the polarization tensor. In quark matter, the Meissner 
mass has been proven to be
a restrictive stability condition at weak coupling: imaginary Meissner masses in the ``gapless 2SC''
\cite{Huang:2004am} and the ``gapless CFL'' \cite{Casalbuoni:2004tb} phases have revealed these phases to be 
unstable (both phases exhibit breached pairing).   We
also calculate the number susceptibility matrix, whose positivity provides a second restrictive stability condition. In the normal phase this quantity 
is identical to the Debye mass but, as we shall discuss in detail, in the super phase
these are two different quantities.   Finally, we also consider the question of global stability (Clogston limit).   We find that there are states which are locally but not globally stable.  Such metastable states could be very interesting from an experimental point of view.    

In most numerical calculations, the ground state was 
found by solving the gap equation in the thermodynamic ensemble with fixed particle numbers for 
both species. In particular, a stable gapless state with one Fermi surface
was predicted \cite{pao}, 
as well as existence of a gapless state with two Fermi surfaces was deduced from 
an effective theory \cite{son}. In our analysis, we do not construct 
the ground state explicitly. Instead we treat chemical potentials and  
the gap fuction as parameters, analyzing the whole parameter space systematically for different stability 
criteria and relating them with the different possible quasiparticle dispersions. We do so
for three-dimensional systems in the main part, Secs.\ \ref{generalresult} -- \ref{results} 
and compare the results with a two-dimensional system in Sec.\ \ref{2D}.
 
When the optimum homogeneous and isotropic state is unstable, the ground state cannot have these properties.  
Plausible candidates for cold atomic systems include phase separation \cite{Bedaque:2003hi} or 
a ``Larkin-Ovchinnikov-Fulde-Ferrell'' (LOFF) state \cite{LOFF}, wherein the 
Cooper pairs carry nonzero total momentum. Whereas a phase-separated state seems unlikely 
in quark matter because of different color and electric charges of the components, 
LOFF phases have been discussed both for quark matter \cite{LOFF1} and cold atom systems 
\cite{sheehy,He:2006wc}.

\section{Formalism and definitions}
\label{formalism}

\subsection{Order parameter, gauge groups, and partition function}
\label{gaugegroup}

We consider a system of two species of nonrelativistic fermions with
Fermi momenta $\mu_1$ and $\mu_2$ and with equal mass $m$. 
One can think of these two species as fermions with spin up and spin 
down. We assume that
there is a pointlike attractive interaction between the two species, giving
rise to the formation of Cooper pairs. 
The structure of the pairing order parameter is
\be \label{order}
\Phi^+ = \Delta\sigma_2 \, ,
\ee
where $\Delta$ is the gap function and $\sigma_2$ is the second Pauli matrix. 
By choosing $\sigma_2$ we only allow for pairing in the singlet channel, i.e., 
between different fermions. (For the effect of induced intra-species pairing see Ref.\ \cite{Bulgac:2006gh}.)
Using the language of spin, this would be the spin-zero channel.
The notation $\Phi^+$ will become clear below where we consider fermion fields 
in Nambu-Gorkov space.

We consider a gauge theory with gauge group $U(1)\times U(1)$. The corresponding gauge fields 
$A_1$ and $A_2$ shall play the role of external fields
which are screened in the superconductor. They should not be confused with the interaction that 
provides the attractive force between the fermions. 
The gauge group is chosen such that it corresponds to the global group associated with 
particle number conservation. Therefore, our results regarding stability of the superconductor 
are valid also for a pure superfluid, i.e., for an analogous  system with uncharged fermions. Moreover, 
the choice of the gauge group enables us to  
study a possible effect of a mixing of the gauge fields on the stability conditions. The
case we consider is the simplest possible to observe this feature. (In quark matter, 
a more complicated gauge group, $SU(3)\times U(1)$, 
where $SU(3)$ is the gauge group of the strong interaction and $U(1)$ the electromagnetic gauge group, 
leads to more complicated mixing patterns of the gauge fields in a color superconductor.)  

The two generators of $U(1)\times U(1)$ are 
\be
T_1\equiv \left(\begin{array}{cc} 1 & 0 \\ 0 & 0\end{array}\right) \,  ,\qquad  
T_2\equiv \left(\begin{array}{cc} 0 & 0 \\ 0 & 1\end{array}\right) \, . 
\ee
This form of the generators implies that the first 
(second) fermion couples exclusively to the first (second) gauge field $A_1$ ($A_2$) with coupling
constants $g_1$ ($g_2$).  
Then, the order parameter (\ref{order}) is invariant under special simultaneous rotations of the 
subgroups. I.e., with $\varphi_1$ and $\varphi_2$ being the phases of the two subgroups $U(1)$,
the order parameter is invariant, $\sum_{a=1,2}\varphi_a(T_a\sigma_2 + \sigma_2 T_a) = 0$, only
if $\varphi_1=-\varphi_2$. Hence the gauge group is spontaneously broken to a $U(1)$ subgroup,
\be
U(1)_{\varphi_1}\times U(1)_{\varphi_2} \to U(1)_{\varphi_1-\varphi_2} \, .
\ee
Therefore, the two original gauge fields mix with each other, giving rise to two
rotated gauge fields in the superconductor. For one of these new gauge fields we expect  
a vanishing Meissner mass, while the orthogonal new gauge field is expected to attain 
nonvanishing Meissner mass via the Anderson-Higgs mechanism.

Having defined the structure of the fermion fields and the gauge groups, we may now 
write down the partition function which, in the subsequent sections, shall serve
as a starting point to define the relevant physical quantities. The partition function is
\be \label{partition}
{\cal Z} = \int {\cal D}A\,e^{S_A}\,Z_f[A] \, .
\ee
The gauge field part $e^{S_A}$ is not relevant for our study and thus does not have to 
be specified. The fermionic part is  
\bea 
Z_f[A] &=& \int {\cal D}\psi^\dag \,{\cal D}{\psi}\,\exp\Bigg(\int_X \Big\{ \psi^\dag
\Big[i\partial_t \non
&&+\, \frac{1}{2m}\left(\nabla-i\G_a{\bf A}_a\right)^2 +\G_aA_a^0 + \mu\Big]\,\psi \non
&& -\, g (\psi^\dag\sigma_2\psi^*)(\psi^T\sigma_2\psi) \Big\}\Bigg)\, ,
\eea
where $\psi$ is the two-component fermion field, each component corresponding to one fermion species,
and $g$ is the coupling constant of the attractive interaction between different fermion
species. 
The chemical potential is given by the matrix $\mu\equiv{\rm diag}(\mu_1,\mu_2)$. 
The integration over time and space is abbreviated by $\int_X$,  
summation over $a=1,2$ is implied and $\G_{1/2}\equiv g_{1/2}T_{1/2}$. We have denoted the spatial part of
the gauge field by the three-vector ${\bf A}$ while the scalar potential is $A^0$. 
Throughout the paper, we work in units of $\hbar = c = k_B=1$. 

\subsection{Propagators, polarization tensor, and screening masses}
\label{propagators}

Next, we introduce the Nambu-Gorkov
field, $\Psi = (\psi,\psi^*)$, consisting of a particle and a hole field, and formally integrate out 
the fermion fields.
This yields in the mean-field approximation with $\Delta=2g\langle\psi^T\sigma_2\psi\rangle$,
\be \label{intout}
{\cal Z} = \int {\cal D}A\,\exp\left[S_A + 
\frac{|\Delta|^2}{4g}-\frac{1}{2}{\rm Tr}\ln({\cal S}^{-1} + {\cal A})\right] \, .
\ee
The term abbreviated by ${\cal A}$ contains the gauge fields and shall be discussed below. 
First we give the explicit form of the full inverse fermion propagator,
\be \label{fullpropinv}
{\cal S}^{-1} \equiv \left(\begin{array}{cc} [G_0^+]^{-1} & \Phi^- \\ \Phi^+ & [G_0^-]^{-1} 
\end{array}\right) \, ,
\ee
where the $\Phi^+$ is the order parameter given by Eq.\ (\ref{order}), $\Phi^-\equiv (\Phi^+)^\dagger$, and
the inverse free fermion propagators are
\be \label{prop1}
[G_0^\pm]^{-1} = i\partial_t \pm \frac{\nabla^2}{2m} \pm \mu \, .
\ee
Note that ${\cal S}^{-1}$ is a $4\times 4$ matrix, two degrees of freedom coming from the Nambu-Gorkov 
structure and explicitly written in Eq.\ (\ref{fullpropinv}), and two degrees of freedom coming
from the two fermion species, implicitly present in Eq.\ (\ref{fullpropinv}) through the $2\times 2$
matrices $[G_0^\pm]^{-1}$ and $\Phi^\pm$. In the following, we need the explicit form of the 
propagator ${\cal S}$ in momentum space. Let us denote its Nambu-Gorkov components as follows,
\be \label{nambuprop}
{\cal S}(K) = \left(\begin{array}{cc} G^+(K) & F^-(K) \\ F^+(K) & G^-(K) \end{array}\right) \, ,
\ee
where we use the shorthand notation $K\equiv(k_0,{\bf k})$ with $k_0 = -i\omega_n$, 
$\omega_n$ being the fermionic Matsubara frequencies.
By inverting Eq.\ (\ref{fullpropinv}) formally, we obtain 
\begin{subequations}
\bea
G^\pm &=& ([G_0^\pm]^{-1}- \Phi^\mp G_0^\mp \Phi^\pm)^{-1} \, , \\ 
F^\pm &=& -G_0^\mp\Phi^\pm G^\pm \, .
\eea
\end{subequations}
From Eq.\ (\ref{prop1}) we conclude that, in momentum space, 
$[G_0^\pm]^{-1} = {\rm diag}\{k_0\mp [k^2/(2m)-\mu_1],k_0\mp [k^2/(2m)-\mu_2]\}$, 
where $k\equiv |{\bf k}|$.
Hence, we find 
\begin{subequations}
\label{propexplicit}
\bea
G^\pm &=&  
\left(\begin{array}{cc} \frac{k_0\pm[k^2/(2m)-\mu_2]}{(k_0\pm\delta\mu)^2-\epsilon_k^2}  & 0 \\ 0 & 
\frac{k_0\pm[k^2/(2m)-\mu_1]}{(k_0\mp\delta\mu)^2-\epsilon_k^2} \end{array}\right) \, , \\
F^\pm &=&  
-i\Delta \left(\begin{array}{cc} 0 & -\frac{1}{(k_0\mp\delta\mu)^2-\epsilon_k^2}
 \\ \frac{1}{(k_0\pm\delta\mu)^2-\epsilon_k^2} & 0 \end{array}\right) \, .
\eea
\end{subequations}
Here and in the following, we use the notation 
\bea
\e_k^2 &\equiv& \xi_k^2+\Delta^2 \, , \qquad
\xi_k\equiv \frac{k^2}{2m}-\bar{\mu} \, , \non 
\delta\mu&\equiv&\frac{\mu_1-\mu_2}{2} \, , \qquad 
\bar{\mu}\equiv \frac{\mu_1+\mu_2}{2} \, . \label{definitions}
\eea
Moreover, we have assumed the gap function $\Delta$ to be real. In the subsequent sections, we shall
make use of these explicit forms of the normal ($G^\pm$) and anomalous ($F^\pm$) propagators.

Let us now come back to Eq.\ (\ref{intout}) in order to define the polarization tensor for the 
gauge fields. In this equation, we have abbreviated
\bea
{\cal A} &\equiv& \left(\begin{array}{cc} A^+ & 0 \\ 0 & A^- \end{array}\right) \, , \non
A^\pm &\equiv& \pm \G_a A_a^0 \mp\frac{\G_a^2}{2m}\,{\bf A}_a^2 \non
&&-\, \frac{i\G_a}{2m}(\nabla\cdot{\bf A}_a+
{\bf A}_a\cdot\nabla) \, .
\eea
We perform a derivative expansion, i.e. expand the logarithm in Eq.\ (\ref{intout}),
and collect the terms which are quadratic in the gauge field.
Let us denote the sum of these terms by $S_2$. After introducing Fourier transforms for 
the fields ${\bf A}_a(X)$, $A^0_a(X)$ and the propagators $S(X,Y)$, and upon assuming translational
invariance, ${\cal S}(X,Y)={\cal S}(X-Y)$, we can write these terms as  
\bea
S_2 &=& -\frac{1}{2}\frac{V}{T}\sum_P \left[ A_a^0(-P)\Pi_{ab}^{00}(P)A_b^0(P) \right.\non 
&& \left.+\,A_a^i(-P)\Pi_{ab}^{i0}(P)A_b^0(P) \right.\non 
&& \left. + \, A_a^0(-P)\Pi_{ab}^{0i}(P)A_b^i(P) \right.\non
&& \left. + \, A_a^i(-P)\Pi_{ab}^{ij}(P)A_b^j(P)\right] \, ,
\eea
where we have defined the following components of the polarization tensor $\Pi$ to one-loop order,
\begin{subequations} \label{polarization}
\be
\Pi_{ab}^{00}(P) \equiv \frac{1}{2}\frac{T}{V}\sum_K {\rm Tr}[S(K)\G_a^-S(K-P)\G_b^-] \, ,
\ee
\bea
\Pi_{ab}^{i0}(P) &\equiv& \frac{1}{2}\frac{T}{V}\sum_K \frac{p_i-2k_i}{2m}\non
&&\times\,{\rm Tr}[S(K)\G_a^+S(K-P)\G_b^-] 
\, ,
\eea
\bea
\Pi_{ab}^{0i}(P) &\equiv& \frac{1}{2}\frac{T}{V}\sum_K \frac{p_i-2k_i}{2m}\non
&&\times\,{\rm Tr}[S(K)\G_a^-S(K-P)\G_b^+] 
\, ,
\eea
\bea
\Pi_{ab}^{ij}(P) &\equiv& \frac{1}{2}\frac{T}{V} \sum_K \left
\{  \frac{\delta_{ij}\delta_{ab}}{m}\,{\rm Tr}[S(K)\bar{\G}_a^2] \right. \non
&&\hspace{-2.5cm}\left. +\frac{(p_i-2k_i)(p_j-2k_j)}{4m^2}  {\rm Tr}[S(K)\G_a^+S(K-P)\G_b^+]\right\} \, . 
\eea
\end{subequations}
The traces run over Nambu-Gorkov and two-fermion space, and we have introduced the
following matrices in Nambu-Gorkov space,
\be
\G_a^\pm \equiv \left(\begin{array}{cc} \G_a & 0 \\ 0 & \pm \G_a  \end{array}\right) \, , \qquad 
\bar{\G}_a^2 \equiv \left(\begin{array}{cc} \G_a^2 & 0 \\ 0 & -\G_a^2 \end{array}\right) \, .
\ee 
In order to compute the electric and magnetic screening masses, we have to compute the 00- and 
$ij$-components of the polarization tensor at zero energy, $p_0=0$ and for vanishing momentum,
${\bf p}\to 0$. The definitions for the Debye and Meissner masses (squared) are 
\begin{subequations}
\bea
m_{D,ab}^2&\equiv& -\lim_{{\bf p}\to 0}\Pi_{ab}^{00}(0,{\bf p}) \, , \\
m_{M,ab}^2&\equiv& \frac{1}{2}\lim_{{\bf p}\to 0}(\delta_{ij}-\hat{p}_i\hat{p}_j)\Pi_{ab}^{ij}(0,{\bf p})\,,  
\eea
\end{subequations}
where $\hat{p}_i\equiv p_i/p$.
With $(\delta_{ij}-\hat{p}_i\hat{p}_j)\delta_{ij} = 2$ and  
$(\delta_{ij}-\hat{p}_i\hat{p}_j)(p_i-2k_i)(p_j-2k_j) = 4k^2[1-(\hat{p}\cdot\hat{k})^2]$ we 
conclude from Eqs.\ (\ref{polarization})
\begin{widetext}
\begin{subequations} \label{debyemeissner}
\bea
\label{debye}
m_{D,ab}^2&=& -\lim_{P\to 0}\frac{1}{2}\frac{T}{V}\sum_K {\rm Tr}[S(K)\G_a^-S(K-P)\G_b^-] \, , \\
\label{meissner}
m_{M,ab}^2&=& \frac{1}{2m}\lim_{P\to 0}\frac{T}{V}\sum_K \left\{\delta_{ab}{\rm Tr}[S(K)\bar{\G}_a^2]  
+\frac{k^2}{2m}\,[1-(\hat{p}\cdot\hat{k})^2]\,{\rm Tr}[S(K)\G_a^+S(K-P)\G_b^+]\right\} \, .
\eea
\end{subequations}
\end{widetext}
These $2\times 2$ matrices in two-fermion space shall be evaluated in the following sections 
in order to obtain stability conditions for gapless superconductors.

\subsection{Pressure, gap equation, and number susceptibilities} 
\label{pressure}

Besides the screening masses, we shall test the number susceptibility matrix $\chi$ on its
positive definiteness. In this section, we define $\chi$ via the thermodynamic pressure
$p$. One may derive the pressure from the partition function in 
(\ref{partition}) using the Cornwall-Jackiw-Tomboulis formalism \cite{cjt}. Employing this 
formalism results in the effective potential, being a functional of the fermion and gauge field
propagators. The pressure is the negative of the effective potential at its stationary 
point (i.e., with the propagators determined to extremize the effective potential). The
fermionic part of the pressure is  
\be \label{pressure01}
p = \frac{1}{2}\frac{T}{V}{\rm Tr}\ln{\cal S}^{-1} + \frac{1}{2}\frac{T}{V}{\rm Tr}[{\cal S}_0^{-1}{\cal S} -1]
+\Gamma_2[{\cal S}] \, , 
\ee
where ${\cal S}_0 = {\rm diag}(G_0^+,G_0^-)$ is the tree-level fermion propagator in Nambu-Gorkov
space with $G_0^\pm$ given in Eq.\ (\ref{prop1}) and $\Gamma_2[{\cal S}]$  is the sum of all
two-particle irreducible diagrams.
The stationarity of the effective potential is ensured by the Schwinger-Dyson 
equation
\be \label{schwingerdyson}
{\cal S}^{-1} = {\cal S}_0^{-1} + \Sigma \, , 
\ee
where
\be \label{selfenergy}
\Sigma\equiv 2\frac{\delta\Gamma_2}{\delta{\cal S}}  
\ee
is the fermion self-energy. 
By making use of the Schwinger-Dyson equation, the pressure can be written as
\be \label{p}
p = \frac{1}{2}\frac{T}{V}{\rm Tr}\ln{\cal S}^{-1}+ \frac{1}{4}\frac{T}{V}{\rm Tr}[{\cal S}_0^{-1}{\cal S} -1]
\, . 
\ee
From Eqs.\ (\ref{pressure01}), (\ref{schwingerdyson}), (\ref{selfenergy}), and the definition of 
the number densities 
\be \label{numberdensity}
n_a = \frac{\partial p}{\partial \mu_a} \, ,
\ee
we conclude
\be \label{numberdens}
n_a = \frac{1}{2}\frac{T}{V}{\rm Tr}\left[{\cal S}\frac{\partial {\cal S}_0^{-1}}{\partial \mu_a}\right]
= \frac{1}{2g_a}\frac{T}{V}\sum_K{\rm Tr}[\Gamma_a^-{\cal S}(K)] \, .
\ee
The number susceptibility $\chi$ is defined as the derivative of the number density with respect to 
the chemical potential (at constant volume and temperature). Hence, making use of
\be
\frac{\partial {\cal S}}{\partial\mu_b} = -{\cal S}\,\frac{\partial {\cal S}^{-1}}{\partial\mu_b}\,
{\cal S}=-{\cal S}\left(\frac{\G_b^-}{g_b}+\frac{\partial \Sigma}{\partial\mu_b}\right)
{\cal S}\, ,
\ee
we obtain 
\bea \label{susc}
\chi_{ab}&=& \frac{\partial n_a}{\partial \mu_b}= -\frac{1}{2g_ag_b}\frac{T}{V}\sum_K{\rm Tr}[\Gamma_a^-{\cal S}(K)\Gamma_b^-{\cal S}(K)]\non
&&-\,\frac{1}{2g_a}\frac{T}{V}\sum_K{\rm Tr}\left[\Gamma_a^-{\cal S}(K)\frac{\partial \Sigma(K)}
{\partial\mu_b}{\cal S}(K)\right] \, .
\eea
The first term on the right-hand side of this equation
is given by the one-loop result for the electric screening mass, cf.\ Eq.\ (\ref{debye}).
For the second term, we assume that the self-energy $\Sigma$ depends on $\mu_b$ only through the 
gap $\Delta$, $\frac{\partial\Sigma}{\partial\mu_b}
=\frac{\partial\Sigma}{\partial\Delta}\frac{\partial\Delta}{\partial\mu_b}$. Then, with 
\bea
-{\cal S}\,
\frac{\partial\Sigma}{\partial\Delta}\,{\cal S}
= -{\cal S}\,
\frac{\partial {\cal S}^{-1}}{\partial\Delta}\,{\cal S}
= \frac{\partial {\cal S}}{\partial\Delta}
\eea
we obtain for the susceptibility  
\be \label{chidef}
\chi_{ab} = \frac{m_{D,ab}^2}{g_ag_b} +
\frac{\partial n_a}{\partial \Delta}\frac{\partial \Delta}{\partial \mu_b} \, .
\ee
In general, the self-energy $\Sigma$
contains terms of any number of fermion loops. Consequently, the number susceptibility
contains terms of arbitrary many fermion loops too, corresponding to
the exact Debye mass including all possible perturbative insertions.
Remarkably, the free fermion result for $\chi$, i.e. $\Sigma=0$, gives the
one-loop result for $m_D^2$ \cite{kapusta}.
We shall use Eq.\ (\ref{chidef}) in the following sections to compute the number susceptibility.
As this equation shows, it goes beyond the one-loop result for the electric screening mass.

One of the off-diagonal components of Eq.\ (\ref{schwingerdyson})
is the self-consistent mean-field gap equation. Using the one-loop approximation for $\Sigma$
we obtain
\be \label{gapequation}
\Phi^+ = -g\frac{T}{V}\sum_K F^+(K) \, ,
\ee
(The other off-diagonal component of Eq.\ (\ref{schwingerdyson}) is simply the hermitian conjugate 
of this equation.) We shall use this gap equation below.

\bigskip
\section{Calculation of screening masses and number susceptibilities}
\label{generalresult}

\subsection{Screening masses}

In this subsection we start from the definitions for the Debye and Meissner masses, 
Eqs.\ (\ref{debyemeissner}), and derive expressions for these masses that shall
be evaluated first in the weak coupling limit, Sec.\ \ref{BCSlimit}, and then for the general
case, Sec.\ \ref{results}. In Eqs.\ (\ref{debye}) and (\ref{meissner}), we first have to 
perform the trace over both Nambu-Gorkov and two-particle space. Then, we perform
the sum over the fermionic Matsubara frequencies, set $p_0=0$, and take the limit
${\bf p}\to 0$. Finally, we take the zero-temperature limit $T\to 0$. Details of this
calculation are deferred to Appendix \ref{matsubara}.  
The results are
\begin{subequations} \label{debyeU1U1}
\bea 
m_{D,11}^2 &=& \frac{g_1^2}{2}\int\frac{d^3{\bf k}}{(2\pi)^3}\Bigg[\frac{\Delta^2}{2\e_k^3}\Theta(\e_k-\d\mu)
\non 
&&+\,\frac{(\e_k+\xi_k)^2}{2\e_k^2}\delta(\e_k-\d\mu)\Bigg] \, , \label{example} 
\\
m_{D,22}^2 &=& \frac{g_2^2}{2}\int\frac{d^3{\bf k}}{(2\pi)^3}\Bigg[\frac{\Delta^2}{2\e_k^3}\Theta(\e_k-\d\mu)
\non
&&+\,\frac{(\e_k-\xi_k)^2}{2\e_k^2}\delta(\e_k-\d\mu)\Bigg] \, , 
\\
m_{D,12}^2 &=& m_{D,21}^2 = \frac{g_1g_2}{2}\non
&& \hspace{-2cm}\times\, \int\frac{d^3{\bf k}}{(2\pi)^3}\left[\frac{\Delta^2}{2\e_k^3}
\Theta(\e_k-\d\mu)-\frac{\Delta^2}{2\e_k^2}\delta(\e_k-\d\mu)\right] \, .
\eea
\end{subequations}
and
\begin{widetext}
\begin{subequations} \label{meissnerU1U1}
\bea
m_{M,11}^2 &=&\frac{g_1^2}{2m}\int\frac{d^3{\bf k}}{(2\pi)^3}\left[\frac{\e_k-\xi_k}{\e_k}+
\frac{\e_k+\xi_k}{\e_k}\Theta(\d\mu-\e_k)\right] \non
&&-\, \frac{g_1^2}{2m}\int\frac{d^3{\bf k}}{(2\pi)^3}
\frac{k^2\sin^2\theta_{\bf k}}{2m} 
\left[\frac{\Delta^2}{2\e_k^3}\Theta(\e_k-\d\mu) + \frac{(\e_k+\xi_k)^2}{2\e_k^2}\delta(\e_k-\d\mu)\right]
\, , \label{m11}
\eea
\be
m_{M,22}^2 =\frac{g_2^2}{2m}\int\frac{d^3{\bf k}}{(2\pi)^3}
\frac{\e_k-\xi_k}{\e_k}\Theta(\e_k-\d\mu)- \frac{g_2^2}{2m}\int\frac{d^3{\bf k}}{(2\pi)^3}
\frac{k^2\sin^2\theta_{\bf k}}{2m} 
\left[\frac{\Delta^2}{2\e_k^3}\Theta(\e_k-\d\mu) + \frac{(\e_k-\xi_k)^2}{2\e_k^2}\delta(\e_k-\d\mu)\right]
\, , \label{m22} 
\ee
\bea
m_{M,12}^2 &=& m_{M,21}^2 = \frac{g_1g_2}{2m}\int\frac{d^3{\bf k}}{(2\pi)^3}\frac{k^2\sin^2\theta_{\bf k}}{2m}
\left[\frac{\Delta^2}{2\e_k^3}
\Theta(\e_k-\d\mu)-\frac{\Delta^2}{2\e_k^2}\delta(\e_k-\d\mu)\right] \, ,
\eea
\end{subequations}
\end{widetext}
where the notation from Eq.\ (\ref{definitions}) has been used. Without loss of generality, we have assumed
$\mu_1\ge\mu_2$, hence $\delta\mu\ge0$. Furthermore, we subtracted the vacuum contribution in the first
integrals on the right-hand sides of Eqs.\ (\ref{m11}) and (\ref{m22}).

We introduce the dimensionless variables
\be
\rho\equiv \frac{\bar{\mu}}{\Delta} \, , \qquad \eta\equiv \frac{\delta\mu}{\Delta}  \, .
\ee
In the following we shall discuss all our results in terms of these variables. It might seem unconventional
to normalize both $\bar{\mu}$ and $\d\m$ with respect to the gap $\Delta$. However, for our 
purpose, this is the most convenient choice. As shall be clear from the following, this choice
enables us to present both different Fermi surface topologies as well as stability criteria
in a single plot, i.e., without having to choose specific values for parameters such as the fermion mass
and the energy gap.
 
\begin{figure*} [ht]
\begin{center}
\hbox{\includegraphics[width=0.33\textwidth]{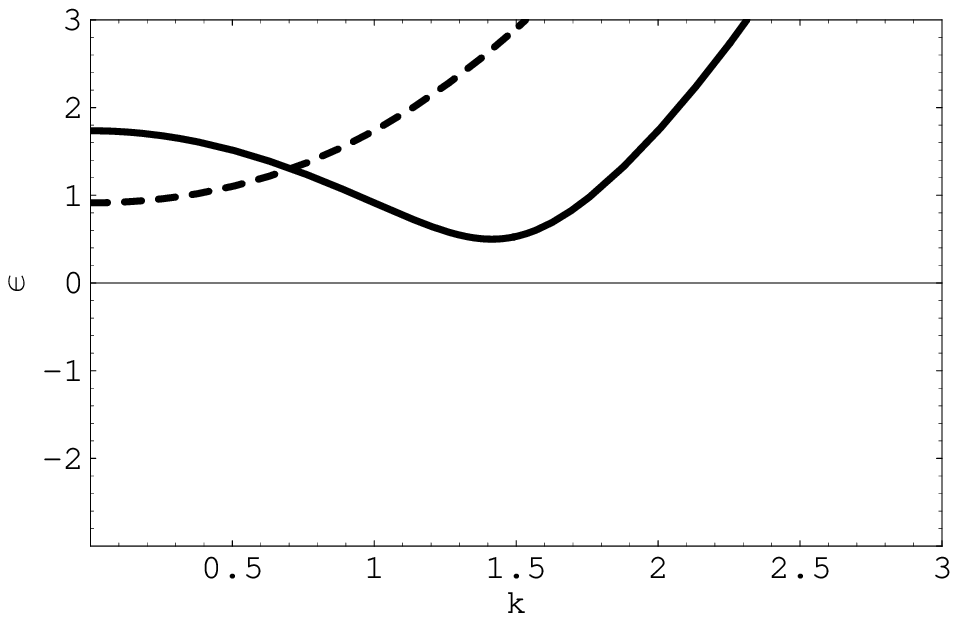}
\includegraphics[width=0.33\textwidth]{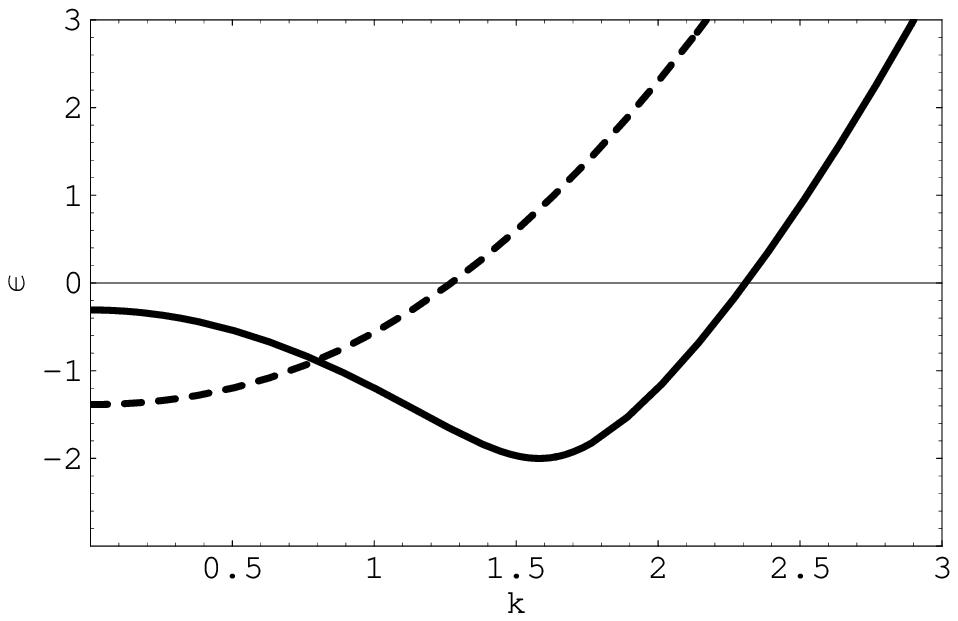}
\includegraphics[width=0.33\textwidth]{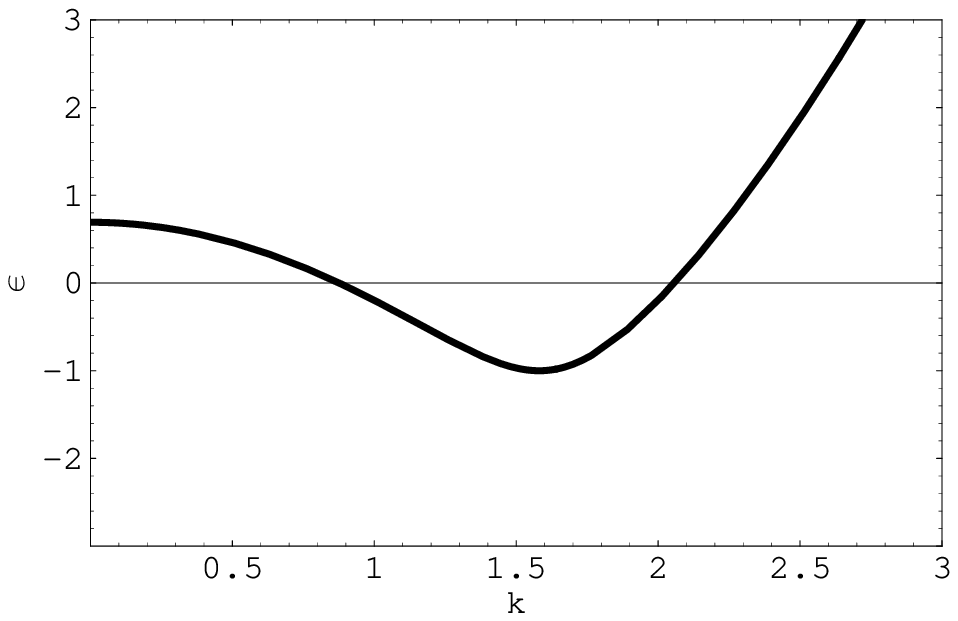}}
\caption{Schematic plot of possible quasiparticle dispersion relations $\e_k^-$ (in arbitrary units).
From left to right, the dispersions correspond to states with zero, one, and two effective Fermi 
surfaces (zeroes of $\e_k^-$). For zero and one effective Fermi surface, two qualitatively 
different dispersions are possible, distinguished by the location of their minimum $k_0$. 
In each case, the solid (dashed) line corresponds to $k_0\neq 0$ ($k_0=0$) and a positive 
(negative) $\bar{\mu}$. From left to right, the parameter ranges are given by Eqs.\ (\ref{F0}), 
(\ref{F1}), and (\ref{F2}),
respectively. }
\label{figdispersions}
\end{center}
\end{figure*}

\begin{figure*} [ht]
\begin{center}
\hbox{\includegraphics[width=0.33\textwidth]{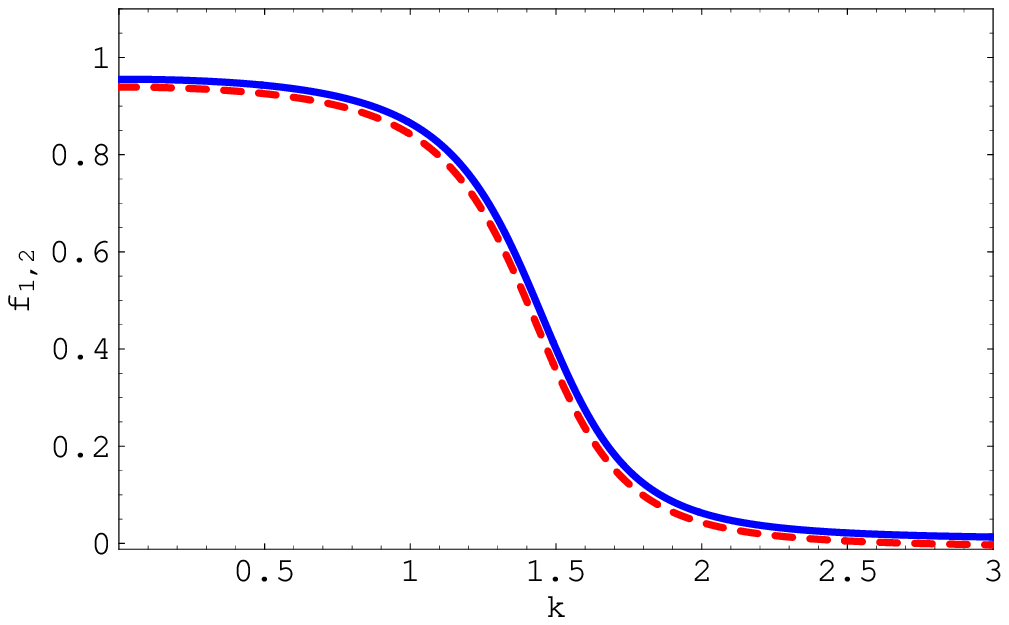}
\includegraphics[width=0.33\textwidth]{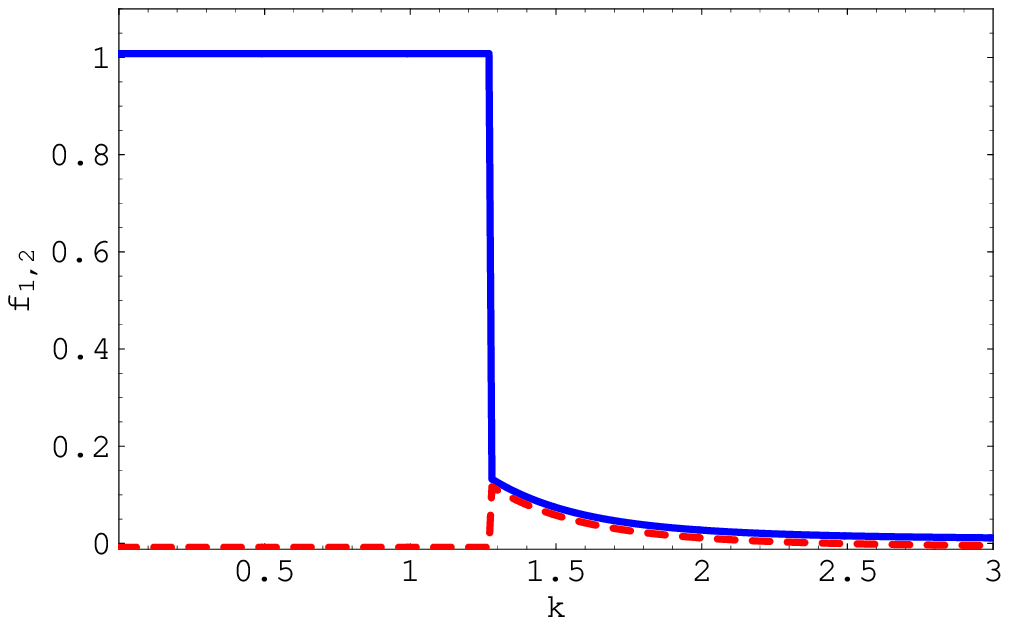}
\includegraphics[width=0.33\textwidth]{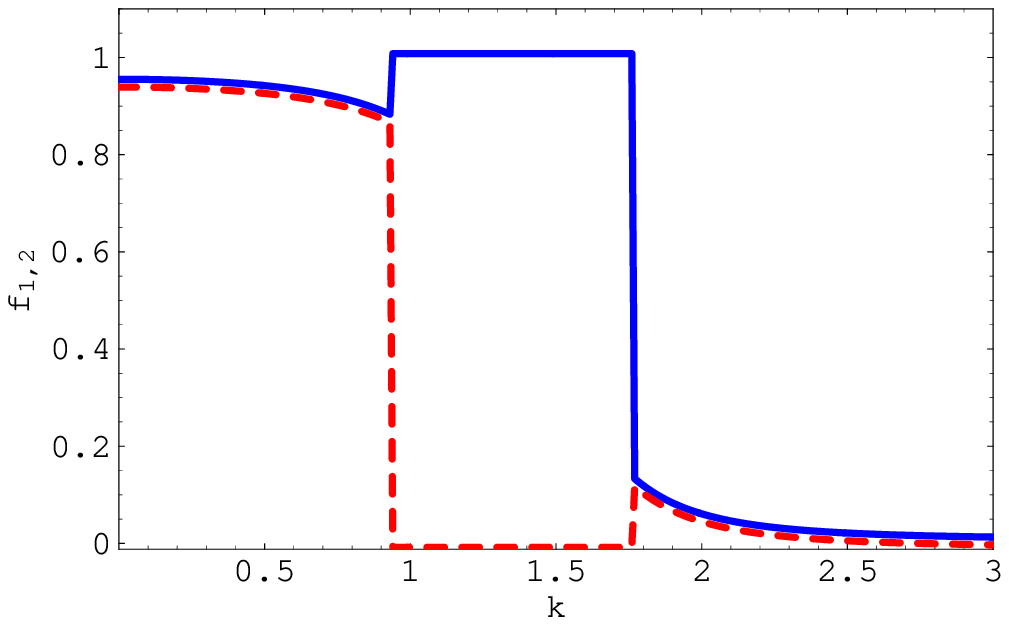}}
\caption{(Color online) Schematic plot of possible quasiparticle occupation numbers (in arbitrary units).
With $\mu_1>\mu_2$, $f_1$ ($f_2$) corresponds to the solid, blue online, (dashed, red online) curves.
From left to right, analogous to Fig.\ \ref{figdispersions}, the occupation numbers correspond 
to states with zero, one, and two effective Fermi surfaces, respectively. (Where both occupation numbers
are identical, e.g., for all $k$ in the left panel, 
we have shifted their value slightly for illustrative purposes.)}
\label{figoccupation}
\end{center}
\end{figure*}

While $\eta\ge 0$, $\rho$ can be both positive or negative, because $\bar{\mu}$ 
assumes negative values in the BEC regime.
Moreover, we abbreviate
\be
\rho_\pm\equiv \rho\pm\sqrt{\eta^2-1} \, , 
\ee
and define the integrals
\begin{subequations} \label{defintegrals}
\bea
\label{defI}
I_\rho(a,b)&\equiv& \int_a^b dx\,\frac{x^2}{[(x^2-\rho)^2+1]^{3/2}} \, ,  \\
\label{defItilde}
\tilde{I}_\rho(a,b)&\equiv& \int_a^b dx\,\frac{x^4}{[(x^2-\rho)^2+1]^{3/2}} \, .
\eea
\end{subequations}
The angular integration in Eqs.\ (\ref{debyeU1U1}) and (\ref{meissnerU1U1}) is trivial. The integration
over the modulus of the fermion momentum $k$ can be done analytically for all terms that contain
the $\delta$-function. To this end, we use the general formula
\bea
\int_0^\infty dk \, k^n\delta(\e_k-\delta\mu)\,f(k) &=& m\,(2m\Delta)^{(n-1)/2} \non
&& \hspace{-4cm}\times\,\Theta(\eta-1)
\frac{\eta}{\sqrt{\eta^2-1}}\left[\Theta(\rho_+)\,\rho_+^{(n-1)/2}\,f(\sqrt{2m\Delta\,\rho_+}) 
\right.\non
&& \hspace{-2cm}\left. + \,\Theta(\rho_-)\,\rho_-^{(n-1)/2}
\,f(\sqrt{2m\Delta\,\rho_-})\right] \label{help1} \, ,
\eea
where $f$ is an arbitrary function, e.g., $f(k)=\Delta^2/(2\e_k^2)$, and $n$ assumes the values $n=2,4$. 
The integrals in Eqs.\ (\ref{debyeU1U1}) and (\ref{meissnerU1U1}) with a $\Theta$-function
lead to elliptic integrals. In general,  
\bea 
\int_0^\infty dk \, k^n \Theta(\e_k-\delta\mu)\,f(k) &=&(2m\Delta)^{(n+1)/2} \non
&& \hspace{-4cm}\times\,\left\{F_n(0,\infty) - \Theta(\eta-1)\left[\Theta(\rho_+)\,F_n(0,\sqrt{\rho_+})
\right.\right.\non
&& \hspace{-2cm} \left.\left.-\,  
\Theta(\rho_-)\,F_n(0,\sqrt{\rho_-})\right]\right\} \, ,\label{help2}
\eea
where again $f$ is an arbitrary function, e.g., $f(k) = \Delta^2/(2\e_k^3)$, and
\be
F_n(a,b)\equiv \int_a^b dx\,x^n f(\sqrt{2m\Delta}\,x) \, .
\ee
The results (\ref{help1}) and (\ref{help2}) show that different terms are switched on or off
by the $\Theta$-functions, depending on whether $\eta$ is larger or smaller than 1 and 
whether $\rho_+$ and $\rho_-$ 
are positive or negative. These conditions can be directly translated into the topology of the 
effective Fermi surfaces, as we demonstrate now. From the denominators in the propagators in 
Eq.\ (\ref{propexplicit}) we conclude that there are two quasiparticle excitation energies,
\be
\e_k^\pm\equiv\sqrt{\left(\frac{k^2}{2m}-\bar{\mu}\right)^2+\Delta^2}\pm\d\m \, .
\ee
It is obvious that, for any values of $\Delta>0$, $\bar{\mu}$ and $\d\m>0$, the first excitation 
branch $\e_k^+$ has no zero. This is 
the usual situation in a conventional superconductor: Quasiparticles at the Fermi surface 
have a finite excitation energy, given by $\Delta$ (and here enhanced by $\d\m$). The second 
quasiparticle dispersion $\e_k^-$, however, is more interesting. Depending on the 
values of $\Delta$, $\bar{\mu}$, and $\d\m$ it can have either no, one, or two zeroes. A zero
of $\e_k^-$ leads to an effective Fermi surface in the superconducting state. Since 
we consider isotropic systems, this surface is a sphere. We illustrate the possible dispersion relations 
in Fig.\ \ref{figdispersions}. In Fig.\ \ref{figoccupation}, we show the corresponding quasiparticle 
occupation numbers. [Their formal expression is encountered in the calculation of the number 
susceptibilities, see integrands on the right-hand sides of Eqs.\ (\ref{densities}).]  
 
\begin{figure}[ht]
\begin{center}
\includegraphics[width=0.5\textwidth]{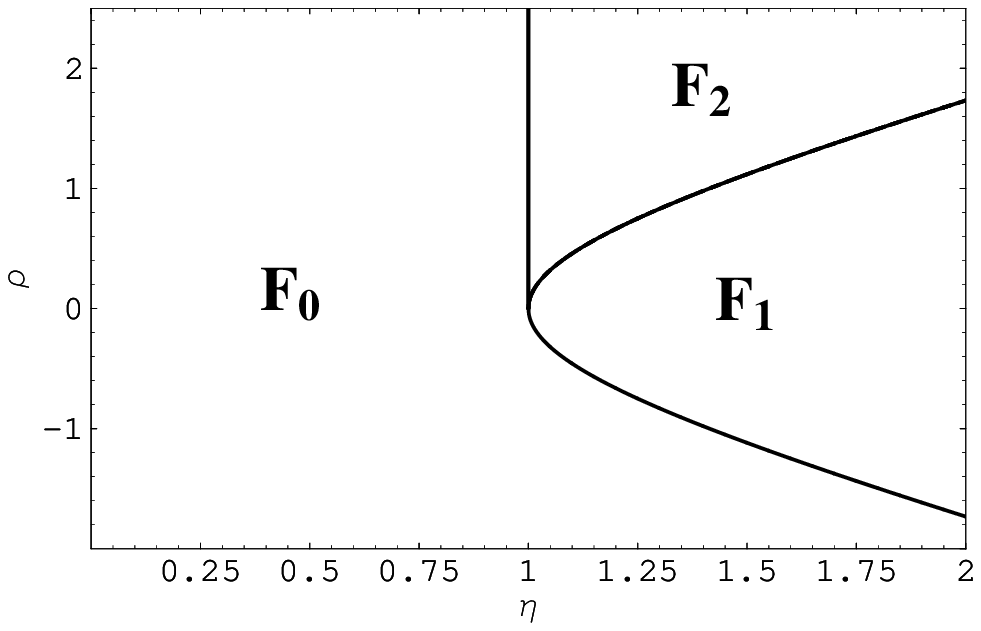}
\caption{Different topologies of the effective Fermi surfaces in the $\rho\eta$-plane, 
$\rho=\bar{\mu}/\Delta$, $\eta=\d\m/\Delta$. Phases in 
the regions $F_0$, $F_1$, and $F_2$ have zero, one, or two effective Fermi surfaces, respectively. Region
$F_1$ is bounded by the lines $\rho_+=0$ and $\rho_-=0$.}
\label{figphases}
\end{center}
\end{figure}

We can classify the different Fermi surface
topologies with the help of the dimensionless parameters:  
\begin{itemize}
\item {\em No effective Fermi surface (region $F_0$ in Fig.\ \ref{figphases})} 
\be \label{F0}
\eta<1 \qquad \mbox{or} \qquad \eta>1,\quad  \rho_+,\rho_-<0 \, .
\ee
The first case, $\eta<1$, translates into $\d\m<\Delta$ and includes the usual BCS 
superconductivity. In the second case, $\eta>1$, $\e_k^-$ has no zeroes provided the   
average chemical potential is sufficiently small, $\bar{\mu}<-\sqrt{\d\m^2-\Delta^2}$. 

\item {\em One effective Fermi surface (region $F_1$ in Fig.\ \ref{figphases})}
\be \label{F1}
\eta>1 , \quad \rho_+>0 , \quad \rho_-< 0 \, .
\ee
This case requires a sufficiently large mismatch of chemical potentials, $\d\m>\Delta$, and 
sets an upper and lower limit for the average chemical potential, 
$-\sqrt{\d\m^2-\Delta^2}<\bar{\mu}<\sqrt{\d\m^2-\Delta^2}$. The latter 
condition cannot be fulfilled in the weak coupling regime. 
The case of one effective Fermi surface is characterized
by unpaired fermions of the first species, occupying the entire effective Fermi sphere 
in momentum space \cite{pao,son}. 

\item {\em Two effective Fermi surfaces (region $F_2$ in Fig.\ \ref{figphases})}
\be \label{F2}
\eta>1, \quad \rho_+,\rho_->0 \, .
\ee
This case is the breached pairing phase \cite{breach}. Here, the average chemical potential has 
a lower bound, $\sqrt{\d\m^2-\Delta^2}<\bar{\mu}$. Unpaired fermions of the first species 
occupy the states between the two Fermi spheres in momentum space. 
\end{itemize}

Having classified the different topologies with respect to the dimensionless quantities, we may now
present the general results for the Debye and Meissner masses. In order to avoid a complicated notation
we may omit all $\Theta$-functions. The rule they impose on the following expressions is simple:
\begin{itemize}
\item For the case of no effective Fermi surface, set $\rho_+=\rho_-=0$.
\item For the case of one effective Fermi surface, set $\rho_-=0$
(and keep $\rho_+$).
\item For the case of two effective Fermi surfaces, keep all terms.
\end{itemize}
These rules are obvious in view of the explicit expressions (\ref{help1}) and (\ref{help2}) 
and the above classifications. In other words, the following expressions are, strictly speaking, only
valid in region $F_2$, but the results for regions $F_0$ and $F_1$ can be obtained immediately 
applying the above rules. 

From now on, for the sake of simplicity, we set the coupling contants $g_1=g_2=1$.  
The result for the Debye mass is then straightforwardly derived as
\begin{subequations} \label{debyerhoeta}
\bea
\frac{m_{D,11/22}^2}{M_D^2} &=& I+\frac{(\eta\pm\sqrt{\eta^2-1})^2}{2\eta\sqrt{\eta^2-1}}\sqrt{\rho_+} \non
&&+\, \frac{(\eta\mp\sqrt{\eta^2-1})^2}{2\eta\sqrt{\eta^2-1}}\sqrt{\rho_-} \, , 
\eea
\bea
\frac{m_{D,12}^2}{M_D^2} &=& \frac{m_{D,21}^2}{M_D^2}= 
I-\frac{\sqrt{\rho_+}+\sqrt{\rho_-}}{2\eta\sqrt{\eta^2-1}} \, . 
\eea
\end{subequations}
The result for the Meissner mass becomes amazingly simple after rewriting some of the integrals.
We show the details of this derivation in Appendix \ref{simplify}. The result is
\bea 
\frac{m_{M,ab}^2}{M_M^2} &=& \tilde{I}-\frac{\rho_+^{3/2}+\rho_-^{3/2}}{2\eta\sqrt{\eta^2-1}} \, . 
\label{meissnerrhoeta}
\eea
We abbreviated
\begin{subequations} 
\bea
I&\equiv& I_\rho(0,\infty)-I_\rho(\sqrt{\rho_-},\sqrt{\rho_+}) \, , \label{definitionI}\\
\tilde{I}&\equiv& \tilde{I}_\rho(0,\infty)
-\tilde{I}_\rho(\sqrt{\rho_-},\sqrt{\rho_+}) \, ,\label{definitionItilde}
\eea
\end{subequations}
and normalized the screening masses with the quantities
\be
M_D^2\equiv \frac{m^{3/2}\Delta^{1/2}}{2\sqrt{2}\pi^2}\, , \quad 
M_M^2\equiv \frac{m^{1/2}\Delta^{3/2}}{3\sqrt{2}\pi^2} \, . 
\ee
Since $M_D^2, M_M^2>0$, and since we are interested in the positive definiteness of the 
screening mass matrices, we can continue our analysis with the 
normalized quantities.

\subsection{Number susceptibilities}

In this subsection we start from the definition of the number susceptibilities (\ref{chidef}) and
compute the matrix $\chi_{ab}$ in order to evaluate it in Secs.\ \ref{BCSlimit} and \ref{results}.
From Eqs.\ (\ref{debyerhoeta}) we already know the first term on the right-hand
side of Eq.\ (\ref{chidef}). The second term is computed as follows. For the derivative 
$\partial n_a/(\partial\mu_b)$ we make use of the fact that the density $n_a$ has basically 
already been computed in the calculation of the Meissner mass, cf.\ Eqs.\ (\ref{meissner}) and
(\ref{numberdens}). Therefore, from Eqs.\ (\ref{m11}) and (\ref{m22}) we read off
\begin{subequations} \label{densities}
\bea
\hspace{-0.5cm}n_1 &=& 
\frac{1}{2}\int\frac{d^3{\bf k}}{(2\pi)^3}\left[\frac{\e_k-\xi_k}{\e_k}+\frac{\e_k+\xi_k}{\e_k}
\Theta(\d\m-\e_k)\right]  , 
\eea
\bea
n_2 &=& \frac{1}{2}\int\frac{d^3{\bf k}}{(2\pi)^3}\frac{\e_k-\xi_k}{\e_k}\Theta(\e_k-\d\m) \, .
\eea
\end{subequations}
Now we straightforwardly take the derivative of the right-hand side of these equations with respect 
to $\Delta$. 
The result, again normalized with the help of $M_D^2$ and written in dimensionless quantities, is
\bea \label{dnddelta}
&&\frac{1}{M_D^2}\frac{\partial n_{1/2}}{\partial\Delta} = 2(\tilde{I} - \rho \,I) \non
&&\hspace{-0.5cm}\mp\,2\left(\frac{\eta\pm\sqrt{\eta^2-1}}{2\eta
\sqrt{\eta^2-1}}\,\sqrt{\rho_+} +\frac{\eta\mp\sqrt{\eta^2-1}}{2\eta\sqrt{\eta^2-1}}\,\sqrt{\rho_-}\right) \, .
\eea
For the derivative $\partial \Delta/(\partial\mu_b)$ we make use of the gap equation 
(\ref{gapequation}). In order to extract the equation for $\Delta$ from this matrix equation, 
we multiply both sides with $\sigma_2$ and take the trace. Then, after performing the Matsubara 
sum and taking the zero-temperature limit, the gap equation reads
\be \label{gapunregular}
-\frac{1}{g} = \int\frac{d^3{\bf k}}{(2\pi)^3}\frac{1}{2\e_k}\Theta(\e_k-\d\m) \, .
\ee
We may rewrite the gap equation in terms of the $s$-wave scattering length $a$ which 
is given by
\be
\frac{m}{4\pi a} = \frac{1}{g} + \frac{1}{V}\sum_{\bf k}\frac{m}{k^2} \, .
\ee
Then, the gap equation becomes  
\be \label{gapregular}
-\frac{m}{4\pi a} = \int\frac{d^3{\bf k}}{(2\pi)^3}\left[\frac{1}{2\e_k}\Theta(\e_k-\d\m)-\frac{m}{k^2}
\right] \, .
\ee
Although not necessary in the current calculation, this rewriting is crucial in order
to solve the gap equation (see Sec.\ \ref{results}) since it provides a natural regularization for the 
otherwise divergent integral (unlike the weak coupling case, where a natural cut-off 
is provided by the Debye frequency). 
We may now use Eq.\ (\ref{gapregular}), or, in this case,
equivalently, Eq.\ (\ref{gapunregular}) and take the derivative with respect to $\mu_b$ on both 
sides of the equation to obtain 
\bea \label{ddeltadmu}
\hspace{-0.5cm}&&\frac{\partial \Delta}{\partial \mu_{1/2}} = \frac{1}{2}
\left(I-\frac{\sqrt{\rho_+}+\sqrt{\rho_-}}{2\eta\sqrt{\eta^2-1}}\right)^{-1}\Bigg[\tilde{I} - \rho \,I 
\non
&&
\hspace{-0.5cm}\mp\,\left(\frac{\eta\pm\sqrt{\eta^2-1}}{2\eta
\sqrt{\eta^2-1}}\,\sqrt{\rho_+} +\frac{\eta\mp\sqrt{\eta^2-1}}{2\eta\sqrt{\eta^2-1}}\,\sqrt{\rho_-}\right)\Bigg]
 \, .
\eea
It is convenient to consider a susceptibility matrix $\tilde{\chi}$ defined in terms of $n \equiv n_1 + n_2$, 
$\d n \equiv n_1 - n_2$,
and $\bar{\mu}$, $\d\m$ rather than $\chi$ which is defined in terms of  $n_1$, $n_2$, $\mu_1$, $\mu_2$.
The relation between $\tilde{\chi}$ and $\chi$ is
\be \label{chitilde}
\tilde{\chi}= Q\,\chi\, Q \, , \qquad Q\equiv\left(\begin{array}{cc}1&1\\1&-1\end{array}\right)
\, ,
\ee
where the entries of $\chi$ are given in Eq.\ (\ref{chidef}).
Inserting the results (\ref{debyerhoeta}), (\ref{dnddelta}), and (\ref{ddeltadmu}) into Eq.\ (\ref{chidef})
and using the transformation (\ref{chitilde}) we obtain
\be \label{omega}
\frac{\tilde{\chi}}{M_D^2} = \frac{4}{R}
\left(\begin{array}{cc}\omega_{11}&\omega_{12}\\\omega_{21}&\omega_{22}\end{array}\right) \, , 
\ee
where 
\be \label{defR}
R\equiv I-\frac{\sqrt{\rho_+}+\sqrt{\rho_-}}{2\eta\sqrt{\eta^2-1}} \, ,
\ee
and 
\begin{subequations} \label{omegaentries}
\bea
\omega_{11}&\equiv& I^2+(\tilde{I}-\rho\,I)^2+I\,\frac{\eta^2-2}{2\eta\sqrt{\eta^2-1}}(\sqrt{\rho_+}+
\sqrt{\rho_-}) \non
&&-\, (\tilde{I}-\rho\,I)\frac{\sqrt{\rho_+}-\sqrt{\rho_-}}{\eta}-\frac{\sqrt{\rho_+}
\sqrt{\rho_-}}{\eta^2} \, , 
\eea
\bea
\omega_{22}&\equiv&I\,\frac{\eta}{2\sqrt{\eta^2-1}}(\sqrt{\rho_+}+\sqrt{\rho_-}) \, , 
\eea
\bea
\omega_{12}&=& \omega_{21}\equiv I\,\frac{\sqrt{\rho_+}-\sqrt{\rho_-}}{2} \non
&&-\,(\tilde{I}-\rho\,I)
\,\frac{\sqrt{\rho_+}+\sqrt{\rho_-}}{2\sqrt{\eta^2-1}} \, .
\eea
\end{subequations}
This result shall be used in the subsequent sections.

\section{Weak coupling limit}
\label{BCSlimit}

Before evaluating the results of the previous section in full generality, let us first 
discuss the weak coupling, or BCS, limit. This limit is characterized by a fixed average 
chemical potential $\bar{\mu}$ which is much larger than both the mismatch $\d\m$ and the 
gap $\Delta$. From this property we conclude that 
the scenario with a single effective Fermi surface is not possible. This scenario is 
only possible in the strong coupling regime. 
However, besides the ordinary superconducting phase without gapless excitation, there is the
possibility of a gapless phase with two effective Fermi surfaces. This situation occurs for
sufficiently large mismatches $\d\m>\Delta$, or $\eta>1$. 

In terms of the above introduced dimensionless parameters, the BCS limit yields the
following simple approximations for the integrals defined in Eq.\ (\ref{defintegrals}),
\begin{subequations}
\bea
I_\rho(0,\infty)&\simeq& \sqrt{\rho} \,\,  , \\
I_\rho(\sqrt{\rho_-},\sqrt{\rho_+})&\simeq& \sqrt{\rho}\,
\frac{\sqrt{\eta^2-1}}{\eta} \, , \\
\tilde{I}_\rho(0,\infty)&\simeq& \rho^{3/2} \,\,  ,  \\
\tilde{I}_\rho(\sqrt{\rho_-},\sqrt{\rho_+})
&\simeq& \rho^{3/2}\,\frac{\sqrt{\eta^2-1}}{\eta} \, .
\eea
\end{subequations}
In all other terms we may approximate 
\be
\rho_-\simeq\rho_+\simeq\rho \, .
\ee
Using these approximations, we can immediately compute explicitly the screening mass 
matrices as well as the number susceptibility matrix from Eqs.\ (\ref{debyerhoeta}), 
(\ref{meissnerrhoeta}), and (\ref{omega}). Remember that in all these equations
we omitted the $\Theta$-functions. In order to compute the BCS limit results, we reinstall the 
factor $\Theta(\eta-1)$ in front of all terms that contain $\rho_-$ or $\rho_+$. Because of
$\bar{\mu}\gg \d\m, \Delta$ we have $\Theta(\rho_+) = \Theta(\rho_-)=1$. 
Then, we obtain for the Debye mass matrix
\begin{subequations}
\bea
\frac{m_{D,11/22}^2}{M_D^2} \simeq \sqrt{\rho}\left(1+\frac{\Theta(\eta-1)\,\eta}{\sqrt{\eta^2-1}}
\right) \, , \\ 
\frac{m_{D,12/21}^2}{M_D^2} \simeq \sqrt{\rho}\left(1-\frac{\Theta(\eta-1)\,\eta}{\sqrt{\eta^2-1}}
\right) \, ,
\eea
\end{subequations}
which leads to the eigenvalues
\be
\frac{m_{D,1}^2}{M_D^2} \simeq 2\sqrt{\rho} \, , \quad 
\frac{m_{D,2}^2}{M_D^2} \simeq 2\sqrt{\rho}\,\frac{\Theta(\eta-1)\,\eta}{\sqrt{\eta^2-1}} \, , 
\ee
which both are positive for all values of the mismatch $\eta$. 

For the Meissner masses, we obtain
\be
\frac{m_{M,ab}^2}{M_M^2}\simeq\rho^{3/2}\left(1-\frac{\Theta(\eta-1)\,\eta}{\sqrt{\eta^2-1}}\right) \, , 
\ee
leading to the eigenvalues
\be
\frac{m_{M,1}^2}{M_M^2} \simeq 2\rho^{3/2}\left(1-\frac{\Theta(\eta-1)\,\eta}{\sqrt{\eta^2-1}}\right)
 \, , \quad 
\frac{m_{M,2}^2}{M_M^2} \simeq 0 \, .
\ee
Hence, in the weak coupling limit, the magnetic screening mass matrix is positive definite if and only if the 
spectrum is fully gapped, $\eta<1$ \cite{He:2005te}. This result is similar to the behavior of the 
gluon Meissner masses
in two- and three-flavor color-superconducting quark matter \cite{Huang:2004am,Casalbuoni:2004tb}.
(However, in a two-flavor color superconductor the situation is more complicated,
because some of the gluon Meissner masses become imaginary even in the fully gapped region. The 
nature of this different kind of instability has been analyzed in 
Ref.\ \cite{Gorbar:2006up}.) 
The result has also been discussed in the context of superfluids, 
where the role of the Meissner mass squared is played by the density of superfluid fermions 
\cite{Gubankova:2004wt,wuyip,zhuang}. 

Finally, for the number susceptibility matrix in the weak coupling limit we find
\begin{subequations}
\bea
\sqrt{\rho}\,R &\simeq&\omega_{11}  \simeq \rho\,\left[1-\frac{\Theta(\eta-1)\,\eta}{\sqrt{\eta^2-1}}\right] \, , \\
\omega_{22} &\simeq& -\rho\,\Theta(\eta-1)\,\left(1-\frac{\eta}{\sqrt{\eta^2-1}}\right) \, , \\
\omega_{12} &=& \omega_{21} \simeq 0 \, , 
\eea
\end{subequations}
and hence 
\be
\frac{\tilde{\chi}}{M_D^2} \simeq 4\sqrt{\rho}\left(\begin{array}{cc} 1& 0 \\ 0 & -\Theta(\eta-1)\end{array}
\right)\, .
\ee
This matrix shows that the gapped state is stable, while the gapless state is unstable, which 
is the same conclusion that can be drawn from the Meissner mass matrix. The negative 
value in $\tilde{\chi}$ corresponds to a negative derivative of $\d n$ with respect to 
$\d \mu$, meaning that, in 
a potential gapless state, an increase in the mismatch of chemical potentials would lead 
to a decrease in the mismatch of number densities. It agrees with physical intuition 
that this cannot be possible.

\section{Stability conditions}
\label{results}

In this section, we use the results from Sec.\ \ref{generalresult} to analyize the stability
of a superconducting state with a given mismatch in chemical potentials and a given 
average chemical potential. This will result in a two-dimensional phase diagram with stable 
and unstable regions. The analysis goes beyond the BCS limit of the previous section by 
allowing for arbitrary values of the mismatch and chemical potentials. In particular, 
we generalize the results to the BEC regime (however, of course, relying still on the mean-field 
results of the previous sections).
Moreover, we solve the gap equation, which we present, for given coupling strengths,
as lines in the phase diagram. 

\subsection{Stability with respect to the screening masses}
\label{stability1}

Both the Debye and Meissner mass matrix are required to be positive definite, i.e., both 
eigenvalues of each of these 2$\times$2 matrices have to be positive. 
The eigenvalues of the 2$\times$2 screening mass matrices $m_I^2$  ($I=D,M$) are given by
\be \label{eigenvals}
m_{I,1/2}^2 = \frac{{\rm Tr}\,m_I^2}{2}\pm\sqrt{\left(\frac{{\rm Tr}\,m_I^2}{2}\right)^2
-{\rm Det}\,m_I^2} \, .
\ee
Since the matrices we consider are real and symmetric, both eigenvalues are real, i.e., 
the argument of the square root in Eq.\ (\ref{eigenvals}) is positive. Hence, $m_{I,1}^2>m_{I,2}^2$, and
both eigenvalues are positive if and only if both ${\rm Tr}\,m_I^2$ and ${\rm Det}\,m_I^2$ are positive. 
We use Eqs.\ (\ref{debyerhoeta}) to compute trace and determinant for the Debye mass matrix. 
We obtain
\begin{subequations}
\bea
\frac{{\rm Tr}\,m_D^2}{M_D^2} &=& 2\,I+ 
\frac{2\eta^2-1}{\eta\sqrt{\eta^2-1}}(\sqrt{\rho_+}+\sqrt{\rho_-}) \, , \label{tracedebye}\\
\frac{{\rm Det}\,{m_D^2}}{M_D^4} &=& I \,
\frac{2\eta}{\sqrt{\eta^2-1}}(\sqrt{\rho_+}+\sqrt{\rho_-}) \non
&&+\, 
4\sqrt{\rho_+}\sqrt{\rho_-}   \, .\label{detdebye}
\eea
\end{subequations}
These expressions can be translated immediately into all three different topologies of the
effective Fermi surfaces, see text above Eq.\ (\ref{debyerhoeta}). In particular, the second term
on the right-hand side of Eq.\ (\ref{tracedebye}) is ``switched on'' only for $\eta>1$. 
Thus this term is positive. Also the integral $I$ is positive, see definitions  
(\ref{defI}) and (\ref{definitionI}). Consequently, both trace and determinant are positive 
for any pair of parameters $\rho$, $\eta$ and hence the Debye mass matrix is positive definite. 

The eigenvalues of the Meissner mass matrix are easily computed from Eq.\ (\ref{meissnerrhoeta}). This
equation shows that all four entries of the matrix are identical.  
Therefore, one of the eigenvalues is zero, while the other one is twice the right-hand side of
Eq.\ (\ref{meissnerrhoeta}). 
This is in accordance with the general argument presented 
in Sec.\ \ref{gaugegroup}. It reflects the unbroken group $U(1)_{\varphi_1-\varphi_2}$. In other words,
there is no Meissner effect for one special admixture of the original gauge fields. The nonvanishing 
second eigenvalue corresponds to the orthogonal admixture. Note that the mixing between the
magnetic part of the gauge fields does not depend on $\rho$ and $\eta$ (it only depends on the coupling 
constants $g_1$, $g_2$, which we have omitted here). This is 
in contrast to the mixing in the electric sector. From the above results for the Debye mass, 
Eqs.\ (\ref{debyerhoeta}), we see that the mixing depends on $\rho$ and $\eta$. 
Only in a fully gapped superconductor, where Eqs.\ (\ref{debyerhoeta}) reduce to 
$m_{D,ab}^2/M_D^2 = I_\rho(0,\infty)$, the mixing
is identical to the magnetic sector. Also in a two-flavor 
color superconductor the mixing in the electric sector depends, in contrast to the magnetic sector,
on the mismatch, shown for the weak coupling limit in Ref.\ \cite{Huang:2004am}. However, in this case, 
different mixing angles of electric and magnetic sectors are observed even in the fully gapped 
phase \cite{Schmitt:2003aa}. 

\begin{figure}[ht]
\begin{center}
\includegraphics[width=0.5\textwidth]{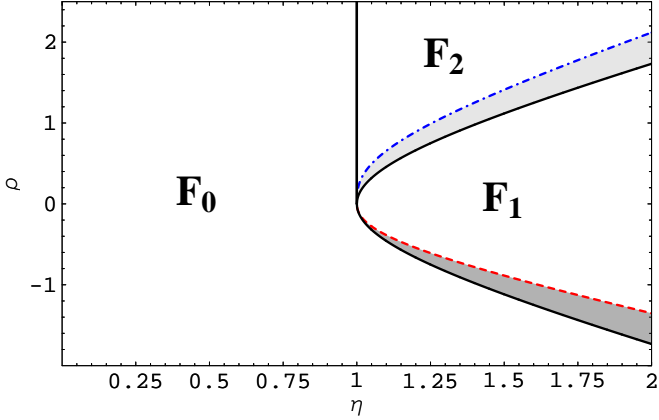}
\caption{(Color online) Stability conditions in the $\rho\eta$-plane. Solid lines and 
regions $F_0$, $F_1$, $F_2$ 
are taken from Fig.\ \ref{figphases}.
The region between the solid vertical line and the dashed-dotted (blue online) curve is unstable 
with repect to a 
negative Meissner mass squared. The region shaded in light gray contains breached pair states that are stable
with respect to the Meissner mass. The region between the solid vertical line and the 
dashed (red online) line is unstable with respect to a negative number susceptibility. The 
region shaded in dark gray contains gapless states with one Fermi sphere that are stable
with respect to the number susceptibility. Remember $\rho=\bar{\mu}/\Delta$, $\eta=\d\m/\Delta$.
}
\label{figinstabilities}
\end{center}
\end{figure}

The stability with respect to the Meissner mass is analyzed by checking the sign of the nonzero
eigenvalue, i.e., the sign of the right-hand side of Eq.\ (\ref{meissnerrhoeta}). 
It is obvious that this eigenvalue is positive for a fully gapped excitation spectrum, 
because $\tilde{I}$ is positive, see definitions (\ref{defItilde}) and (\ref{definitionItilde}).
It can be checked numerically that the eigenvalue is also positive in the case 
of one Fermi surface, i.e., setting $\rho_-=0$ in Eq.\ (\ref{meissnerrhoeta}). 
The most interesting case is the breached phase. 
In this case there is a region in the parameter space where the eigenvalue is negative, indicating
a magnetic instability. This region is separated from the stable region by a line in 
the $\rho\eta$-plane, given by the (numerical) solution to the equation
\be \label{meissnerzero}
\tilde{I}_\rho(0,\infty)-\tilde{I}_\rho(\sqrt{\rho_-},\sqrt{\rho_+})
-\frac{\rho_+^{3/2}+\rho_-^{3/2}}{2\eta\sqrt{\eta^2-1}}=0   \, .   
\ee
The solution is given by the dashed-dotted (blue online) curve in Fig.\ \ref{figinstabilities}. It renders
all phases between the black vertical and the dashed-dotted line unstable with 
respect to the Meissner mass. However, this stability criterion leaves 
an apparently stable breached pair region in the phase diagram (shaded in light gray).

\subsection{Stability with respect to the number susceptibility}
\label{stability2}

In this subsection we test the susceptibility matrix (\ref{omega}) on its positive definiteness.
From the transformation (\ref{chitilde}) it is obvious that $\tilde{\chi}$ is 
positive definite if and only if $\chi$ is positive definit (note that $Q^2=2$). 
So we may restrict our analysis 
to $\tilde{\chi}$. Again, the simplest case is the one without zero in the dispersion relations.
In this case, there is only one nonzero entry in the susceptibility matrix. This entry is positive for all
$\rho$ (and independent of $\eta$),
\begin{subequations}
\bea
\frac{\tilde{\chi}_{11}}{M_D^2} &=& 4\,\frac{I_\rho^2(0,\infty)+
[\tilde{I}_\rho(0,\infty)-\rho I_\rho(0,\infty)]^2}{I_\rho(0,\infty)} \, , \\
\tilde{\chi}_{12}&=&\tilde{\chi}_{21}=\tilde{\chi}_{22}=0 \, .
\eea
\end{subequations}
The most convenient way to determine the sign of the eigenvalues of $\tilde{\chi}$ in the 
other two cases is to write them as
\be
\frac{\tilde{\chi}_{1/2}}{M_D^2} = \frac{4}{R}\,
\left[\frac{{\rm Tr}\,\omega}{2}\pm\sqrt{\left(\frac{{\rm Tr}\,\omega}{2}\right)^2-{\rm Det}\,
\omega}\right]\, .
\ee
where the entries of the $2\times 2$ matrix $\omega$ are given by Eqs.\ (\ref{omegaentries}), and $R$ is defined in 
Eq.\ (\ref{defR}). 
For two effective Fermi surfaces, it is
a simple numerical task to determine the signs of ${\rm Det}\, \omega$ and
$R$. Both quantities turn out to be negative
 throughout the parameter region of interest. Consequently, independent of the sign of ${\rm Tr}\, \omega$,
at least one eigenvalue is negative, rendering the complete parameter region 
of two effective Fermi surfaces unstable. 

For one effective Fermi surface, one can show numerically that ${\rm Tr}\, \omega$ is positive
throughout the parameter region. The determinant can be written as
\bea \label{detomega}
{\rm Det} \, \omega &=&  
R\,\frac{\eta\sqrt{\rho_+}}{2\sqrt{\eta^2-1}}\left\{I_\rho^2(\sqrt{\rho_+},\infty) \right.\non
&&\left.+\, 
[\tilde{I}_\rho(\sqrt{\rho_+},\infty) - \rho\,I_\rho(\sqrt{\rho_+},\infty)]^2\right\} \, .
\eea
This expression is very useful because it enables us to find a simple condition for the 
positivity of $\tilde{\chi}$ without computing the eigenvalues explicitly. From 
Eq.\ (\ref{detomega}) it is clear that ${\rm Det}\,\omega$ is positive if and only if $R$ is positive.
Therefore, the positivity of $\tilde{\chi}$ reduces to the positivity of $R$:
If $R<0$, one of the eigenvalues of $\tilde{\chi}$ is negative, while for $R>0$
both eigenvalues are positive. Hence, using the definition for $R$, Eq.\ (\ref{defR}), 
stable and unstable regions in the $\rho\eta$-plane are separated by the solution of the equation 
\be  \label{line2} 
I_\rho(\sqrt{\rho_+},\infty)-\frac{\sqrt{\rho_+}}{2\eta\sqrt{\eta^2-1}} = 0 \, .
\ee
Interestingly, the structure of this equation is very similar to that of 
Eq.\ (\ref{meissnerzero}). The dashed (red online) line
in Fig.\ \ref{figinstabilities} represents the solutions to Eq.\ (\ref{line2}). Consequently, 
the region between the solid vertical line and the dashed curve is unstable. 
This result shows that 
the number susceptibility alone would have been sufficient to determine stable and unstable regions
of the phase diagram in our approach. It is interesting to see that neither of the stability conditions 
exactly coincides with the topology of the dispersion relation. In other words, the lines that separate 
stable from unstable regions are close to, but not on top of the lines that separate regions
of different topology. The conclusion in terms of topology is that the entire region 
without an effective Fermi surface is stable, not a single point in the region with two effective
Fermi surfaces (``breach'') is stable, and a subset of the region with one effective Fermi surface
is stable. This subset only contains phases with negative
average chemical potential. We recall that, within our assumption of a homogeneous system,
we are not able to make predictions about the true ground states in the unstable region.
In the context of cold atoms, it can be expected that a negative Meissner mass squared leads
to a LOFF state \cite{sheehy,He:2006wc,Mannarelli:2006hr}, while a negative number susceptibility 
leads to phase separation \cite{Bedaque:2003hi,sheehy}.
    
At the end of this section, we note that, in the considered model, 
the positivity of the number susceptibility corresponds to a local maximum of the pressure. 
This can be seen explicitly as follows. 
We start from Eq.\ (\ref{p}), insert the propagators, Eqs.\ (\ref{nambuprop}) 
and (\ref{propexplicit}), take the trace, perform the Matsubara sum, and take the 
zero-temperature limit. This yields
\be \label{pressure1}
p=\int\frac{d^3{\bf k}}{(2\pi)^3}\left[\left(\e_k-\frac{\Delta^2}{2\e_k}\right)\Theta(\e_k-\delta\mu)
+\delta\mu\,\Theta(\delta\mu-\e_k)\right] \, .
\ee
We find that the second derivative of the pressure with respect to the gap is proportional to the 
function $R$ that also appears in the number susceptibility,
\be
\frac{\partial^2p}{\partial\Delta^2} = - 4 M_D^2\, R \, .
\ee  
Consequently, from the above discussion of the eigenvalues $\tilde{\chi}_1$, $\tilde{\chi}_2$
we conclude
\be \label{equivalence}
\frac{\partial^2p}{\partial\Delta^2} < 0 \quad \Leftrightarrow \quad \tilde{\chi}_1, \tilde{\chi}_2 >0 \, .
\ee

\subsection{Solutions to the gap equation}
\label{gapsolutions}

\begin{figure}[ht]
\begin{center}
\includegraphics[width=0.5\textwidth]{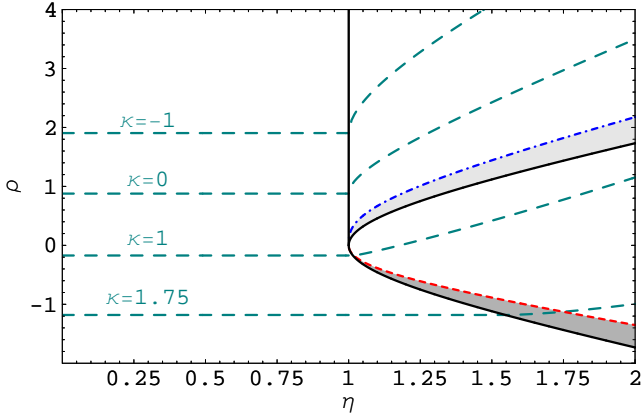}
\caption{(Color online) The solid lines are taken from Fig.\ \ref{figphases}. The dashed-dotted (blue online) 
and short-dashed (red online) curves as well as the shaded areas are taken from Fig.\ \ref{figinstabilities}.
The long-dashed (green online) curves are solutions of the gap equation for different coupling 
strengths $\kappa$.}
\label{figgapeq}
\end{center}
\end{figure}

The stability analyis in the previous two subsections was performed without knowledge of the
explicit solution of the gap equation: The screening masses are completely independent 
of the gap equation. For the number susceptibility, we made use of the gap equation 
in order to extract the derivative of the gap with respect to the two chemical
potentials (which was done without solving the gap equation). Consequently, the main results of 
this paper, regions of (in)stability of the homogeneous phases and their connection to
the regions of different Fermi surface topologies, do not need the explicit form of the 
gap function. However, one might say that not all regions of the phase diagram are 
accessible for solutions of the gap equation. Moreover, one would like to know
which regions of the phase diagram are accessible to which coupling strengths. 
Therefore, in this subsection, we present 
the solution of the gap equation and its representation as lines in the phase diagram. 
We do so by first rewriting the gap equation (\ref{gapregular}) in the above used dimensionless 
quantities, 
\be \label{gaptau}
-\kappa = K_\rho^\prime(0,\infty) - K_\rho(\sqrt{\rho_-},\sqrt{\rho_+}) \, , 
\ee
where
\begin{subequations} 
\bea  
&&\hspace{-0.5cm}K_\rho^\prime(0,\infty) \equiv \int_0^\infty dx\,
\left[\frac{x^2}{\sqrt{(x^2-\rho)^2+1}}-1\right] \, , \\
&&\hspace{-0.5cm}K_\rho(\sqrt{\rho_-},\sqrt{\rho_+})\equiv \int_{\sqrt{\rho_-}}^{\sqrt{\rho_+}} 
dx\,\frac{x^2}{\sqrt{(x^2-\rho)^2+1}} \, .
\eea
\end{subequations}
We have introduced the new dimensionless parameter
\be
\kappa \equiv \frac{\pi}{2\sqrt{2m\Delta}\,a } \, , 
\ee
which plays the role of the coupling strength: Small values, $\kappa\to-\infty$, correspond to the 
BCS limit, whereas 
large values, $\kappa\to +\infty$, correspond to the BEC limit. We solve Eq.\ (\ref{gaptau})      
for a fixed $\kappa$ to obtain $\rho$ as a function of $\eta$. Let us explain this in somewhat more detail:
After fixing $\kappa$, 
we first consider the region with no effective Fermi surface, i.e., we set $\rho_-=\rho_+=0$. In this
case, the right-hand side of Eq.\ (\ref{gaptau}) does not depend on $\eta$, i.e., we obtain a constant
number, say $\rho_0(\kappa)$, represented by 
a horizontal line in the phase diagram, see Fig.\ \ref{figgapeq}. Physically, this means that the 
average chemical potential does not change upon increasing the mismatch. This is true until 
the mismatch $\eta$ reaches a critical value and the constant line reaches an area in the
phase diagram with gapless excitations. Determined by the value $\rho_0(\kappa)$, this area can either 
be the region
of one or of two effective Fermi surfaces. (Or it can be the ``splitting point'' shown in the middle of the 
figure, which is a special case that corresponds to the point ``S'' in the phase diagram in Ref.\ \cite{son}.
This point is hit by the line with $\kappa\simeq 0.84$). To obtain the continuation of the 
constant line, we switch on 
the terms $\rho_+$ (for one Fermi surface) or both $\rho_-$ and $\rho_+$ (for two Fermi surfaces) 
in Eq.\ (\ref{gaptau}) and 
solve the equation from the point $\eta=\sqrt{\rho_0^2(\kappa)+1}$ (for one Fermi surface) or
$\eta=1$ (for two Fermi surfaces). In Fig.\ \ref{figgapeq} we show the solutions for four different
coupling strengths, $\kappa = -1,0,1,1.75$, two of which end up in the breached region 
for sufficiently large mismatches and two of which end up in the region with a single effective 
Fermi sphere. In particular, the line $\kappa=0$, corresponding to the Feshbach resonance 
$a=\pm\infty$, hits the breached pair region. This is in agreement with previous 
mean-field studies.

All lines enter an unstable region for sufficiently large mismatches, meaning that an inhomogeneous 
phase, e.g., a LOFF or mixed phase, or the normal state take over.  
For all couplings $\kappa$ that result in $\rho_0(\kappa)<0$, the system passes, 
for mismatches 
$\eta>\sqrt{\rho_0^2(\kappa)+1}$, through a stable gapless region with one effective Fermi surface
and reaches, for even larger mismatches ($\eta$ such that the left-hand side 
of Eq.\ (\ref{line2}) is smaller than zero), an unstable region.

\section{Generalization of Clogston limit}

So far we have investigated the stability of (gapless) superconductors with respect to a 
real magnetic screening mass and a positive number susceptibility. 
We have seen that the latter condition is equivalent to a local maximum of the pressure,
Eq.\ (\ref{equivalence}). 
In this section,  
we ask which of the states in the phase diagram cannot be {\em global} maxima of the pressure.
To this end, we compare the pressure in the superconducting state $p_s$ with the one in the
normal conducting state $p_n$ and require the difference to be positive, $p_s-p_n>0$.
We do so for fixed chemical potentials. Therefore, $p_n$ is the pressure of the vacuum
in the case of negative chemical potentials. It is expected that 
the same analysis for fixed number densities leads to a different result.
  
\begin{figure} [ht]
\begin{center}
\includegraphics[width=0.5\textwidth]{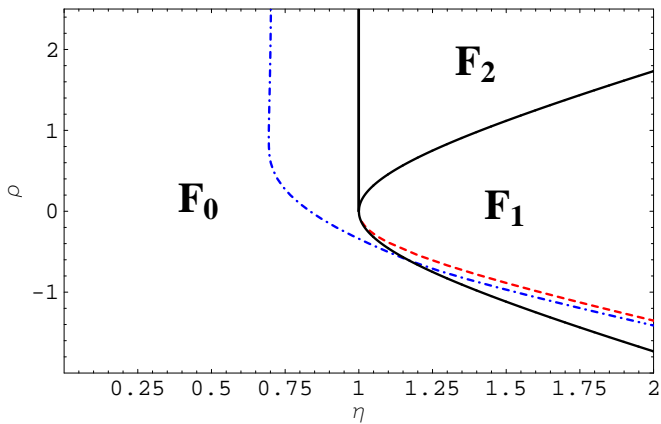}
\caption{(Color online) The solid lines are taken from Fig.\ \ref{figphases}. 
Dashed-dotted (blue online) line: Solution to $\Delta p=0$. For all states above (below)  this line, 
$\Delta p<0$ ($\Delta p>0$). For comparison, we also show the line given by the susceptibility
condition (dashed, red online), see Fig.\ \ref{figinstabilities}.}
\label{figglobal}
\end{center}
\end{figure}

The pressure in the normal phase is obtained by setting $\Delta=0$ in Eq.\ (\ref{pressure1}), 
\be
\frac{p_n}{p_0} = \frac{\lambda^5}{5}-\rho\frac{\lambda^3}{3} + \frac{2}{15}[\Theta(\rho+\eta)
(\rho+\eta)^{5/2}+\Theta(\rho-\eta)(\rho-\eta)^{5/2}] \, ,
\ee
where, in order to use the same dimensionless parameters as in the previous section, we introduced 
the gap $\Delta$ as the energy scale. Then, the pressure is given in units of  
\be
p_0\equiv\frac{\sqrt{2}m^{3/2}\Delta^{5/2}}{\pi^2} \, , 
\ee
and the cut-off $\Lambda$ for the momentum integral in Eq.\ (\ref{pressure1}) is replaced by the 
dimensionless cut-off $\lambda \equiv \Lambda/\sqrt{2m\Delta}$. After writing the pressure $p_s$ for
the superconducting phase in dimensionless quantities, we find for the 
pressure difference the cut-off-independent expression 
\bea \label{deltap}
\Delta p&\equiv& \frac{p_s-p_n}{p_0} = L^\prime_\rho(0,\infty)-L_\rho(\sqrt{\rho_-},\sqrt{\rho_+}) \non
&&\hspace{-1cm}
+\,\frac{4}{15}\rho^{5/2} + \frac{\eta}{3}
(\rho_+^{3/2}-\rho_-^{3/2}) \non
&&\hspace{-1cm}-\,\frac{2}{15}[\Theta(\rho+\eta)
(\rho+\eta)^{5/2}+\Theta(\rho-\eta)(\rho-\eta)^{5/2}] \, . 
\eea
Here we have, as in the previous sections, omitted the $\Theta$-functions that come with $\rho_+$ and
$\rho_-$, and we have abbreviated the integrals
\begin{subequations}
\bea  
L^\prime_\rho(0,\infty) &\equiv& \int_0^\infty dx\,x^2\Bigg[\sqrt{(x^2-\rho)^2+1} \non
&& \hspace{-1cm} -\,\frac{1}{2\sqrt{(x^2-\rho)^2+1}}-|x^2-\rho|\Bigg] \, ,\\
L_\rho(\sqrt{\rho_-},\sqrt{\rho_+}) &\equiv& \int_{\sqrt{\rho_-}}^{\sqrt{\rho_+}}dx\,x^2
\Bigg[\sqrt{(x^2-\rho)^2+1} \non
&&-\,\frac{1}{2\sqrt{(x^2-\rho)^2+1}}\Bigg] \, .
\eea
\end{subequations}
In Fig.\ \ref{figglobal} we show the 
solution to the equation $\Delta p=0$, represented by the dashed-dotted (blue online) line. All states
above (and right from) this line cannot be global maxima of the pressure because $\Delta p<0$. 
We observe that the 
region excluded by this condition contains both the region of negative number susceptibility and imaginary
Meissner mass. However, a gapless region with a single Fermi surface and $\Delta p>0$ remains.
Extending the results to larger mismatches than shown in the figure, $\eta>2$, we find that the dashed and dashed-dotted lines approach each other, that is, for large 
$\eta$ the stability conditions given by $\tilde{\chi}$ and $\Delta p$ come to coincide. 

The condition $\Delta p>0$ excludes some fully gapped states. In particular, one may
extend the results to large chemical potentials, corresponding to the weak coupling regime. 
One finds that the dashed-dotted line in Fig.\ \ref{figglobal} approaches a vertical line given by
$\eta\simeq 1/\sqrt{2}$. That is no surprise, since here the original discussion by Clogston applies \cite{clogston}. Analytically, in the weak coupling limit, 
\be   
L^\prime_\rho(0,\infty) \simeq \frac{\sqrt{\rho}}{4} \, .
\ee
Then, setting $\rho_-=\rho_+=0$ in Eq.\ (\ref{deltap}) and expanding the expression for the pressure in the 
normal phase for $\rho\gg\eta$, we obtain
\be
\Delta p\simeq \frac{\sqrt{\rho}}{2}\left(\frac{1}{2}-\eta^2\right) \, ,
\ee
and we see that this expression is indeed negative for $\eta>1/\sqrt{2}$.

\section{Stability in two dimensions}
\label{2D}

In this section, we re-do our stability calculations for $d=2$ dimensional systems. 
This case has the advantage that the stability conditions can be discussed in forms
of simple analytical expressions, as we show below.
Apart from their theoretical convenience, one can certainly imagine performing experiments with 
quasi-two-dimensional superfluids/superconductors with mismatched Fermi surfaces.

We notice that Fig.\ \ref{figphases} only depends on the form of the dispersion, $\e_k\sim k^2$. 
Therefore, the boundaries of the regions $F_0$, $F_1$, and $F_2$ 
in this figure are valid also for two dimensions.
The results for the Meissner mass and the number susceptibility, however, depend on the dimension.
For $d=2$ we use Eq.\ (\ref{meissnerrhoeta}) and (\ref{omega}) -- (\ref{omegaentries})
with the following replacements. The elliptic integrals in Eqs.\ (\ref{defintegrals}) become simple analytic 
expressions after reducing the powers of $x$ in the numerator of the integrands by 1,
\begin{subequations} 
\bea
I_\rho^{2d}(a,b)&=& \left. \frac{1}{2}\frac{x^2-\rho}{\sqrt{(x^2-\rho)^2+1}}\right|_{x=a}^{x=b}  \, ,  
\\
\tilde{I}_\rho^{2d}(a,b)&=& \left. \frac{1}{2}\frac{\rho(x^2-\rho)-1}{\sqrt{(x^2-\rho)^2+1}}\right|_{x=a}^{x=b} \, .
\eea
\end{subequations}
The integration boundaries do not depend on $d$, i.e., after evaluating the $\Theta$-function, 
$\sqrt{\rho_+}$ and $\sqrt{\rho_-}$ appear as boundaries also in two dimensions. 
Again, the reason is that these quantities only depend on the dispersion, cf.\ Eq.\ (\ref{help2}).
In the remaining terms, the powers of $\rho_+$ and $\rho_-$ change. From Eqs.\ (\ref{help1}),
(\ref{help2}), and the results in Appendix \ref{simplify}, we conclude that in the expressions for the 
Meissner mass squared, $\rho_\pm^{3/2}$ has to be replaced by $\rho_\pm$. 
It should be mentioned that our definition for the Meissner mass cannot naively be
transfered to two dimensions (simply using a two-dimensional volume $V$ in Eq.\ (\ref{meissner}) would 
not yield dimensions $[{\rm energy}]^2$ for $m_M^2$). However, this is no problem for our 
stability analysis.
The simplest way to think of the Meissner mass squared is as follows. In our model, the global number 
conservation group is spontaneously broken in the same way as the gauge group. Therefore, 
the Meissner mass squared is proportional to the number density of superfluid particles. 
The latter quantity has dimensions $[{\rm energy}]^d$, i.e., it depends on the dimension $d$.  
Thus, although we are, strictly speaking, not discussing the Meissner mass in two dimensions, we can use
the same stability condition (\ref{meissnerzero}) with the above replacements for $d=2$
(and refer to it in the rest of this section as ``stability with respect to the Meissner mass''). 
For the breached pair phase, Eq.\ (\ref{meissnerzero}) becomes 
\be
\frac{1}{2}(\rho+\sqrt{\rho^2 +1})-\frac{\rho\eta}{\sqrt{\eta^2-1}} = 0 \,\, .
\ee
This equations has no solution in the relevant parameter range, i.e., for $\rho>\sqrt{\eta^2-1}$. The 
left-hand side is negative for all relevant $\eta$, $\rho$. Consequently, with respect to the Meissner
mass, there is no stable breached pair state. For 
the case of a single effective Fermi sphere, this equation reads
\be
\rho\left(1-\frac{\eta}{\sqrt{\eta^2-1}}\right)=0 \,\, .
\ee
This equation is only solved by $\rho = 0$. Because the expression in parentheses is negative, states 
with $\rho>0$ are unstable while states with $\rho<0$ are stable with respect to the Meissner mass. 
As for $d=3$ it is obvious that all fully gapped states are stable.

In the expression for the number susceptibility, $\sqrt{\rho_\pm}$ has to be replaced
by $\rho_\pm^0 = 1$. For the breached pair state, the expressions for the eigenvalues of 
$\tilde{\chi}$ are, even for $d=2$, 
too lengthy to show here. One checks numerically that at least one of the eigenvalues is negative.
The phases with a single Fermi surface
are treated with the same analysis as described in Sec.\ \ref{stability2}. Hence,
we may use the two-dimensional analogue of Eq.\ (\ref{line2}) which is
\be
1-\frac{\eta}{\sqrt{\eta^2-1}}=0 \,\, .
\ee
This equation has no solution. Since the left-hand side is negative for any $\eta$, all 
states with a single Fermi sphere are unstable for $d=2$. 

\begin{figure} [ht]
\begin{center}
\includegraphics[width=0.5\textwidth]{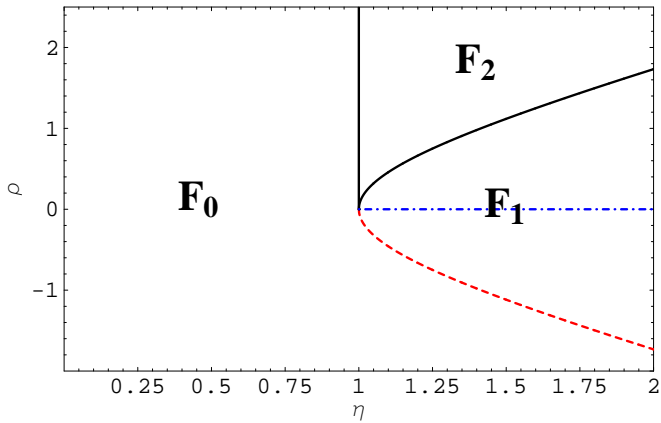}
\caption{(Color online) Unstable regions in two-dimensional systems. The region between the solid vertical and the 
dashed (red online) lines is unstable with respect to the number susceptibility. The region between the 
solid vertical and the dashed-dotted (blue online) lines is unstable with respect to the Meissner mass.
The dashed line exactly coincides with the line that separates fully gapped phases from phases 
with one effective Fermi surface.}
\label{fig2D}
\end{center}
\end{figure}

We summarize these results in Fig.\ \ref{fig2D}. This figure should be compared to its analogue in 
three dimensions, Fig.\ \ref{figinstabilities}. We observe that both stability conditions 
are more restrictive in two dimensions. The condition of the 
positivity of the Meissner mass squared does not leave any stable breached pair phase. And
the positivity of the number susceptibility does not leave any stable phase with 
one effective Fermi surface. Again, the unstable region
with respect to the Meissner mass is a subset of the unstable region with respect to the number 
susceptibility. So the stable regions are determined solely by the number susceptibility. 
Unlike in the case $d=3$, the lines that separate stable from unstable regions are exactly
on top of lines that separate different Fermi surface topologies.  
In some recent papers general connections between stability and Fermi surface topology 
have been proposed \cite{Horava:2005jt,Volovik:2006gt}. General results for topological 
stability are obtained using K-theory \cite{Horava:2005jt} and are given in terms of the dimensionality 
of the system and of the effective Fermi surface. It would be interesting to see if similar analyses 
apply to the situations considered in this paper.   

\section{Conclusions}
\label{conclusions}

We have studied stability conditions and their relation to different Fermi surface topologies in a 
superconductor of two fermion species with mismatched chemical potentials. The results have been 
presented in a phase diagram in which the mismatch and the average chemical potential serve as parameters.
Determining the zeroes of one of the two quasiparticle excitation energies provides lines in the
phase diagram that separate regions with zero, one and two effective Fermi spheres (these 
lines are given by simple analytic expressions). 
We have computed the Meissner mass (squared) and the number susceptibility matrices. Both of these 
$2\times 2$ matrices have to be positive definite in a stable state. Consequently, they define
lines in the phase diagram separating stable from unstable phases with respect to these conditions. 
Up to remaining elliptic integrals, we have derived analytic expressions for these lines. 
The two expressions are amazingly similar. Moreover they are close to, but not identical 
with, the lines that separate phases with different Fermi surface topologies. We have found that 
the condition with respect to a real Meissner mass leaves a stable breached pair phase. However, 
this region is rendered unstable by a negative eigenvalue of the number susceptibility matrix. 
Stable gapless states with a single Fermi surface exist for negative average chemical potentials.
Solution of the gap equation shows that these states are on the BEC side of the 
Feshbach resonance.

Comparing local and global stability conditions, we have identified possible regions of metastability.  That possibility seems particularly interesting from an experimental point of view, because the onset of catastrophic and practically unpredictable (nucleated) decay, as a function of coupling constant or number density, would be a striking physical phenomenon.  

Our results suggest several future studies. Interesting theoretical 
questions are related to the apparent connection of the stability conditions with the Fermi surface
topology.  Is there a qualitative reason why the stability lines lie very close to, but do not coincide with, the lines marking topology change?
Is there a qualitative reason why phases with 
a single Fermi sphere are stable (at least in a certain region of the phase diagram), but phases
with two effective Fermi spheres are unstable? From the experimental point of view, it remains challenging 
to define practically accessible signatures of the gapless phases. Our calculation of the screening masses 
could be used to predict transport properties of atomic gapless superfluids.

\begin{acknowledgments}
The authors thank M.\ Alford, A.\ Chitov, M.\ Mannarelli, E.\ Mishchenko, I.\ Shovkovy, and M.\ Zwierlein
for valuable comments and discussions. 
AS acknowledges support by the German Academic Exchange 
Service (DAAD) and the U.S. Department of Energy
under contracts DE-FG02-91ER50628 and DE-FG01-04ER0225 (OJI).  FW acknowledges support from the U. S. Department of Energy contract DE-FG02-05ER41360. 
\end{acknowledgments}

\appendix 

\section{Matsubara sums for the screening masses}
\label{matsubara}

In this appendix, we explain how to obtain the expressions (\ref{debyeU1U1}) and (\ref{meissnerU1U1})
from the definitions (\ref{debyemeissner}). In particular, this calculation includes
the summation over fermionic Matsubara frequencies. 
For the Debye and Meissner masses we need to compute the following quantities,
\bea
M_a&\equiv& T\sum_{n}{\rm Tr}[S(K)\bar{\G}_a^2] \non
&=& g^2_a\, T\sum_{n}{\rm Tr}[G^+(K)T_a-G^-(K)T_a] \, , \label{Mab}
\eea
and
\bea
N_{ab}^\pm &\equiv& T\sum_{n}{\rm Tr}[S(K)\G_a^\pm S(K-P)\G_b^\pm]\non
&=& g_a g_b\,T\sum_{n}{\rm Tr}[G^+(K)T_aG^+(K-P)T_b \non
&&+\, G^-(K)T_aG^-(K-P)T_b] \non
&& \pm \, T\sum_{n} {\rm Tr}[F^-(K)T_aF^+(K-P)T_b \non
&&+ \,F^+(K)T_aF^-(K-P)T_b] \, , \label{Nab}
\eea
where the trace over Nambu-Gorkov space has been performed.
In order to compute $M_{ab}$, we insert the normal and anomalous propagators from 
Eq.\ (\ref{propexplicit}) into Eq.\ (\ref{Mab}) and perform the trace over two-fermion space. 
In this case, the Matsubara sum can be easily performed. 
We set the gauge boson energy to zero, $p_0=0$ and take the zero-temperature limit 
(where we assume, without loss of generality, $\mu_1>\mu_2$ and hence $\d\mu>0$).
Hereafter, we take the limit of a vanishing gauge boson momentum, ${\bf p}\to 0$ and obtain
\be\label{M1122}
\frac{M_{1/2}}{g_{1/2}^2}=\pm\frac{\e_k\pm\xi_k}{\e_k}\Theta(\d\mu-\e_k) - \frac{\xi_k}{\e_k} \, .
\ee
In order to compute $N_{ab}^\pm$, we insert the fermion propagators  
into Eq.\ (\ref{Nab}). In this case, the Matsubara sums are more complicated. We make use of the 
following two generic formulas, which can be derived via contour integration in the complex $k_0$-plane. 
For real numbers $\a$, $\b$, $\d_1$, $\d_2$, $a>0$, $b>0$, we have 
\begin{widetext}
\bea \label{matsu1}
&& T\sum_n\frac{(k_0+\a)\,(k_0-p_0+\b)}{[(k_0\pm\d_1)^2-a^2]\,[(k_0-p_0\pm\d_2)^2-b^2]}  \non
&=&\frac{1}{4ab}\left\{\frac{(a_\mp+\a)\,(b_\pm-\b)}{p_0-a_\mp-b_\pm}\,[1-n(a_\mp)-n(b_\pm)]
-\frac{(a_\pm-\a)\,(b_\mp+\b)}{p_0+a_\pm+b_\mp}\,[1-n(a_\pm)-n(b_\mp)] \right. \non
&& \left .+ \frac{(a_\pm-\a)\,(b_\pm-\b)}{p_0+a_\pm-b_\pm}\,[n(a_\pm)-n(b_\pm)]
- \frac{(a_\mp+\a)\,(b_\mp+\b)}{p_0-a_\mp+b_\mp}\,[n(a_\mp)-n(b_\mp)] \right\}
\eea
where $k_0=-i(2n+1)\pi T$ and $p_0=-i2m\pi T$ are the fermionic and bosonic Matsubara frequencies, 
respectively, and $n(x)\equiv 1/(\exp(x/T)+1)$ is the Fermi distribution.
Moreover, $a_\pm\equiv a\pm\d_1$, $b_\pm\equiv b\pm\d_2$. And
\bea \label{matsu2}
&&T\sum_n\frac{1}{[(k_0\pm\d_1)^2-a^2]\,[(k_0-p_0\pm\d_2)^2-b^2]} \non
&&=\frac{1}{4ab}\left\{\frac{1}{p_0+a_\pm+b_\mp} \,[1-n(a_\pm)-n(b_\mp)]  
-\frac{1}{p_0-a_\mp-b_\pm}\,[1-n(a_\mp)-n(b_\pm)]  \right.\non
&& \left. + \frac{1}{p_0+a_\pm-b_\pm} \, [n(a_\pm)-n(b_\pm)] -
\frac{1}{p_0-a_\mp+b_\mp} \,[n(a_\mp)-n(b_\mp)]\right\} \, .
\eea
\end{widetext}
After applying these results, we again set $p_0=0$ and take the limits
$T\to 0$, ${\bf p}\to 0$. We obtain
\begin{subequations} \label{N1122}
\be
\frac{N_{11}^\pm}{g_1^2} = -\frac{\Delta^2}{2\e_k^3}\Theta(\e_k-\d\mu)-\frac{(\e_k+\xi_k)^2}{2\e_k^2}\delta(\e_k-\d\mu)
\, ,
\ee 
\be
\frac{N_{22}^\pm}{g_2^2} = - \frac{\Delta^2}{2\e_k^3}\Theta(\e_k-\d\mu)
-\frac{(\e_k-\xi_k)^2}{2\e_k^2}\delta(\e_k-\d\mu)
\, , 
\ee
\bea
&&\hspace{-1cm}\frac{N_{12}^\pm}{g_1g_2} = \frac{N_{21}^\pm}{g_1g_2} \non
&&= \pm\left[\frac{\Delta^2}{2\e_k^3}\Theta(\e_k-\d\mu)
-\frac{\Delta^2}{2\e_k^2}\delta(\e_k-\d\mu)\right] \, . 
\eea
\end{subequations} 
Inserting Eqs.\ (\ref{M1122}) and (\ref{N1122}) into (\ref{debyemeissner}) yields 
(\ref{debyeU1U1}) and (\ref{meissnerU1U1}). 
 
\section{Integrals in Meissner mass}
\label{simplify}

Here we derive the Meissner mass, Eq.\ (\ref{meissnerrhoeta}), from Eqs.\ (\ref{meissnerU1U1}).
With 
\be
J_\rho^\pm(a,b)\equiv\int_a^b dx\,x^2\left(1\pm\frac{x^2-\rho}{\sqrt{(x^2-\rho)^2+1}}\right) \,
\ee
and the integral $\tilde{I}_\rho(a,b)$ defined in Eq.\ (\ref{defItilde}), we obtain from 
Eqs.\ (\ref{meissnerU1U1})  
\begin{widetext}
\begin{subequations}
\bea \label{diagonal}
\frac{m^2_{M,11/22}}{g_{1/2}^2 M_M^2} &=& 3\left[J_\rho^-(0,\infty) \pm J_\rho^\pm(\sqrt{\rho_-},\sqrt{\rho_+})\right]
-\left[\tilde{I}_\rho(0,\infty) - \tilde{I}_\rho(\sqrt{\rho_-},\sqrt{\rho_+})\right] \non
&&-\, 
\left[\frac{(\eta\pm\sqrt{\eta^2-1})^2}{2\eta\sqrt{\eta^2-1}}\,\rho_+^{3/2} +
\frac{(\eta\mp\sqrt{\eta^2-1})^2}{2\eta\sqrt{\eta^2-1}}\,\rho_-^{3/2}\right] \, , 
\eea
\bea
\frac{m^2_{M,12}}{g_1g_2 M_M^2} &=& \frac{m^2_{M,21}}{g_1g_2 M_M^2} = 
\tilde{I}_\rho(0,\infty) - \tilde{I}_\rho(\sqrt{\rho_-},\sqrt{\rho_+}) 
- \frac{\rho_+^{3/2}+\rho_-^{3/2}}{2\eta\sqrt{\eta^2-1}} \, .
\eea
\end{subequations}
The result for the diagonal elements $m_{M,11/22}^2$ is simplified by observing that the integrals 
$J^\pm_\rho$ and $\tilde{I}_\rho$ are in fact related via partial integration,
\bea \label{partialint}
J_\rho^\pm(a,b) &=& \frac{b^3}{3}\left(1\pm\frac{b^2-\rho}{\sqrt{(b^2-\rho)^2+1}}\right) 
- \frac{a^3}{3}\left(1\pm\frac{a^2-\rho}{\sqrt{(a^2-\rho)^2+1}}\right) \mp \frac{2}{3}\tilde{I}_\rho(a,b) \, .
\eea
\end{widetext}
In particular, 
\be \label{partialint1}
J_\rho^-(0,\infty) = \frac{2}{3}\tilde{I}_\rho(0,\infty) \,\, .
\ee
Rewriting $J_\rho^-(0,\infty)$ and $J_\rho^\pm(\sqrt{\rho_-},\sqrt{\rho_+})$ in Eq.\ (\ref{diagonal}) with 
the help of these relations renders the diagonal elements identical to the off-diagonal ones, and
we arrive at Eq.\ (\ref{meissnerrhoeta}).


\begin{thebibliography}{99}

\bibitem{bcs} 
J.\ Bardeen, L.N.\ Cooper, and J.R.\ Schrieffer, 
Phys.\ Rev.\ {\bf 108}, 1175 (1957).

\bibitem{helium3}
A.J.\ Leggett, Rev.\ Mod.\ Phys.\ {\bf 47}, 331 (1975); 
D.\ Vollhardt and P.\ W\"olfle, 
{\it The Superfluid Phases of Helium 3} (Taylor \& Francis, London, 1990).  

\bibitem{highTc}
J.G.\ Bednorz and K.A.\ M\"uller, Z.\ Physik B {\bf 64}, 189 (1986); 
M.\ Sigrist and K.\ Ueda, Rev.\ Mod.\ Phys.\ {\bf 63}, 239 (1991).

\bibitem{spin1}
T.~Sch\"afer,
Phys.\ Rev.\ D {\bf 62}, 094007 (2000)
[arXiv:hep-ph/0006034];
  A.~Schmitt,
  Phys.\ Rev.\ D {\bf 71}, 054016 (2005)
  [arXiv:nucl-th/0412033].

\bibitem{pao}
C.-H.\ Pao, S.-T.\ Wu, S.-K.\ Yip, arXiv:cond-mat/0506437.

\bibitem{breach}
  E.~Gubankova, W.~V.~Liu and F.~Wilczek,
  Phys.\ Rev.\ Lett.\  {\bf 91}, 032001 (2003)
  [arXiv:hep-ph/0304016].


\bibitem{Liu:2002gi}
  W.~V.~Liu and F.~Wilczek,
  Phys.\ Rev.\ Lett.\  {\bf 90}, 047002 (2003)
  [arXiv:cond-mat/0208052].



\bibitem{Gubankova:2004wt}
  E.~Gubankova, F.~Wilczek and E.~G.~Mishchenko,
  Phys.\ Rev.\ Lett.\  {\bf 94}, 110402 (2005)
  [arXiv:cond-mat/0409088].

\bibitem{exp}
C.A.\ Regal, M.\ Greiner, and D.S.\ Jin,
Phys.\ Rev.\ Lett.\ {\bf 92}, 040403 (2004) [arXiv:cond-mat/0401554];
M.\ Bartenstein {\em et. al.}, 
Phys.\ Rev.\ Lett.\ {\bf 92}, 120401 (2004)
[arXiv:cond-mat/0401109];
M.W.\ Zwierlein {\em et. al.}, 
Phys.\ Rev.\ Lett.\ {\bf 92}, 120403 (2004)
[arXiv:cond-mat/0403049];
J.\ Kinast {\em et. al.}, 
Phys.\ Rev.\ Lett.\ {\bf 92}, 150402 (2004)
[arXiv:cond-mat/0403540];
T.\ Bourdel {\em et. al.}, 
Phys.\ Rev.\ Lett.\ {\bf 93}, 050401 (2004)
[arXiv:cond-mat/0403091].

\bibitem{BECBCS}
P.\ Nozieres and S.\ Schmitt-Rink, J.\ Low.\ Temp.\ Phys.\ {\bf 59}, 195 (1985);
M.\ Marini, F.\ Pistolesi, and G.C.\ Strinati,
Eur.\ J.\ Phys.\ B {\bf 1}, 151 (1998)
[arXiv:cond-mat/9703160];
  E.~Babaev,
  Phys.\ Rev.\ B {\bf 63}, 184514 (2001)
  [arXiv:cond-mat/0010085].



\bibitem{KetterleImbalancedSpin}
M.~W.~Zwierlein, A.~Schirotzek, C.~H.~Schunck and W.~Ketterle, 
Science {\bf 311}, 492 (2006) [arXiv:cond-mat/0511197].

\bibitem{HuletPhaseSeparation}
G.~B.~Partridge, W. Li, R.~I.~Kamar, Y.~Liao and R.~G.~Hulet, 
Science {\bf 311}, 503 (2006)
[arXiv:cond-mat/0511752].

\bibitem{Bedaque:2003hi}
  P.~F.~Bedaque, H.~Caldas and G.~Rupak,
  Phys.\ Rev.\ Lett.\  {\bf 91}, 247002 (2003)
  [arXiv:cond-mat/0306694];
  M.~M.~Forbes, E.~Gubankova, W.~V.~Liu and F.~Wilczek,
  Phys.\ Rev.\ Lett.\  {\bf 94}, 017001 (2005)
  [arXiv:hep-ph/0405059];
  J.~Carlson and S.~Reddy,
  Phys.\ Rev.\ Lett.\  {\bf 95}, 060401 (2005)
  [arXiv:cond-mat/0503256].

\bibitem{Bailin:1983bm}
  D.~Bailin and A.~Love,
  Phys.\ Rept.\  {\bf 107}, 325 (1984).

\bibitem{reviews}
For reviews, see
K.~Rajagopal and F.~Wilczek,
arXiv:hep-ph/0011333;
M.~G.~Alford,
Ann.\ Rev.\ Nucl.\ Part.\ Sci.\  {\bf 51}, 131 (2001)
[arXiv:hep-ph/0102047];
D.~H.~Rischke, Prog.\ Part.\ Nucl.\ Phys.\ {\bf 52}, 197 (2004) 
[arXiv:nucl-th/0305030];
I.~A.~Shovkovy,
  Found.\ Phys.\  {\bf 35}, 1309 (2005)
  [arXiv:nucl-th/0410091]. 

\bibitem{Alford:1998mk}
M.~G.~Alford, K.~Rajagopal and F.~Wilczek,
Nucl.\ Phys.\ B {\bf 537}, 443 (1999)
[arXiv:hep-ph/9804403].

\bibitem{Rajagopal:2005dg}
  K.~Rajagopal and A.~Schmitt,
  Phys.\ Rev.\ D {\bf 73}, 045003 (2006)
  [arXiv:hep-ph/0512043].


\bibitem{igor}
  M.~Kitazawa, D.~H.~Rischke and I.~A.~Shovkovy,
  arXiv:hep-ph/0602065.

\bibitem{Huang:2004am}
  M.~Huang and I.~A.~Shovkovy,
  Phys.\ Rev.\ D {\bf 70}, 094030 (2004)
  [arXiv:hep-ph/0408268].

\bibitem{Casalbuoni:2004tb}
  R.~Casalbuoni, R.~Gatto, M.~Mannarelli, G.~Nardulli and M.~Ruggieri,
  Phys.\ Lett.\ B {\bf 605}, 362 (2005)
  [Erratum-ibid.\ B {\bf 615}, 297 (2005)]
  [arXiv:hep-ph/0410401];
  K.~Fukushima,
  Phys.\ Rev.\ D {\bf 72}, 074002 (2005)
  [arXiv:hep-ph/0506080].



\bibitem{son}
  D.~T.~Son and M.~A.~Stephanov,
  arXiv:cond-mat/0507586.






\bibitem{LOFF}
A.~I.~Larkin and Yu.~N.~Ovchinnikov, Zh. Eksp. Teor. Fiz.~{\bf 47}, 1136
(1964)[Sov. Phys. JETP {\bf 20}, 762 (1965)];
P.~Fulde and R.~A.~Ferrell, Phys.\ Rev.\ {\bf 135}, A550 (1964).

\bibitem{LOFF1}
M.~G.~Alford, J.~A.~Bowers and K.~Rajagopal,
Phys.\ Rev.\ D {\bf 63}, 074016 (2001)
[arXiv:hep-ph/0008208];
  I.~Giannakis and H.~C.~Ren,
  Phys.\ Lett.\ B {\bf 611}, 137 (2005)
  [arXiv:hep-ph/0412015];

\bibitem{sheehy}
D.E.\ Sheehy and L.\ Radzihovsky, Phys.\ Rev.\ Lett.\ {\bf 96}, 060401 (2006) [arXiv:cond-mat/0508430].

\bibitem{He:2006wc}
  L.~He, M.~Jin and P.~f.~Zhuang,
  arXiv:cond-mat/0601147.

\bibitem{Bulgac:2006gh}
  A.~Bulgac, M.~McNeil Forbes and A.~Schwenk,
  arXiv:cond-mat/0602274.

\bibitem{cjt}
J.M.\ Cornwall, R.\ Jackiw, and E.\ Tomboulis, Phys.\ Rev.\ D {\bf 10}, 2428 (1974).

\bibitem{kapusta}
J.I.\ Kapusta, {\em Finite-temperature field theory} (Cambridge University Press, Cambridge, 1989). 



\bibitem{shovkovy}
  I.~Shovkovy and M.~Huang,
  Phys.\ Lett.\ B {\bf 564}, 205 (2003)
  [arXiv:hep-ph/0302142].

\bibitem{He:2005te}
  L.~n.~He, M.~Jin and P.~f.~Zhuang,
  Phys.\ Rev.\ B {\bf 73}, 024511 (2005)
  [arXiv:hep-ph/0509317].

\bibitem{Gorbar:2006up}
  E.~V.~Gorbar, M.~Hashimoto, V.~A.~Miransky and I.~A.~Shovkovy,
  arXiv:hep-ph/0602251.

\bibitem{wuyip}
S.-T.\ Wu and S.-K.\ Yip, 
Phys.\ Rev. A {\bf 67}, 053603 (2003) [arXiv:cond-mat/0303185].

\bibitem{zhuang}
L.~n.~He, M.~Jin and P.~f.~Zhuang,
arXiv:cond-mat/0604137. 

\bibitem{Schmitt:2003aa}
  A.~Schmitt, Q.~Wang and D.~H.~Rischke,
  Phys.\ Rev.\ D {\bf 69}, 094017 (2004)
  [arXiv:nucl-th/0311006].

\bibitem{Mannarelli:2006hr}
  M.~Mannarelli, G.~Nardulli and M.~Ruggieri,
  arXiv:cond-mat/0604579.

\bibitem{clogston}
A.M.\ Clogston, Phys.\ Rev.\ Lett.\ {\bf 9}, 266 (1962); B.S.\ Chandrasekhar, Appl.\ Phys.\ Lett.\ {\bf 1}, 7
(1962).

\bibitem{Horava:2005jt}
  P.~Horava,
  Phys.\ Rev.\ Lett.\  {\bf 95}, 016404 (2005)
  [arXiv:hep-th/0503006].

\bibitem{Volovik:2006gt}
  G.~E.~Volovik,
  arXiv:cond-mat/0601372.



\end{thebibliography}
\end{document}